\documentclass[fleqn]{aa}

\usepackage{newtxtext,newtxmath}

\usepackage[T1]{fontenc}

\DeclareRobustCommand{\VAN}[3]{#2}
\let\VANthebibliography\thebibliography
\def\thebibliography{\DeclareRobustCommand{\VAN}[3]{##3}\VANthebibliography}

\usepackage{graphicx}	% Including figure files
\usepackage{amsmath}	% Advanced maths commands
\usepackage{url}
\usepackage[none]{hyphenat}
\usepackage[noend]{algorithmic}
\usepackage{algorithm}
\usepackage{bm}
\usepackage[table]{xcolor}
\makeatletter
\let\linenumbers\relax
\let\runninglinenumbers\relax
\let\setrunninglinenumbers\relax
\let\switchlinenumbers\relax

\let\modulolinenumbers\relax
\let\internallinenumbers\relax
\let\resetlinenumber\relax
\makeatother
\newcommand{\mat}[1]{\bm{\mathrm{#1}}}
\begin{document}
\titlerunning{Interpretable Dark Matter Clustering}

%\title{Breaking Self-Interacting Dark Matter Degeneracies using Interpretable Neural Networks}
\title{Measuring the Dark Matter Self-Interaction Cross-Section with Deep Compact Clustering for Robust Machine Learning Inference}

\author{Ethan Tregidga \inst{1}\thanks{ethan.tregidga@epfl.ch}
          \and David Harvey \inst{1} 
          \and Luca Biggio \inst{2}
          \and Felix Vecchi \inst{1}
          }

\institute{
Laboratoire d’Astrophysique, EPFL, Observatoire de Sauverny, 1290 Versoix, Switzerland
\and Department of Computing Sciences, Bocconi University, Milano, Italy
% \and Lorentz Institute for Theoretical Physics, Leiden University, PO Box 9506, NL-2300 RA Leiden, The Netherlands
% \and Leiden Observatory, Leiden University, PO Box 9513, NL-2300 RA Leiden, The Netherlands
}

% Abstract of the paper
\abstract{

We have developed a machine learning algorithm capable of detecting ``out-of-domain data'' for trustworthy cosmological inference.
By using data from two separate suites of cosmological simulations, we show that our algorithm is able to determine whether ``observed'' data is consistent with its training domain, returning confidence estimates as well as accurate parameter estimations.
We apply our algorithm to two-dimensional images of galaxy clusters from the BAHAMAS-SIDM and DARKSKIES simulations with the aim to measure the self-interaction cross-section of dark matter.
Through deep compact clustering we construct an informative latent space where galaxy clusters are mapped to the latent space forming ``latent-clusters'' for each simulation, with the location of the latent-cluster corresponding to the macroscopic parameters, such as the cross-section, $\sigma_{\rm DM}/m$.
We then pass through mock observations, where the location of the observed latent-cluster informs us of which properties are shared with the training data.
If the observed latent-cluster shares no similarities with latent-clusters from the known simulations, we can conclude that our simulations do not represent the observations and discard any parameter estimations, thus providing us with a method to measure machine learning confidence.
This method serves as a blueprint for transparent and robust inference that is in demand in scientific machine learning.
The code is available via \protect\url{https://github.com/EthanTreg/Bayesian-DARKSKIES}.

}

%%%%%%%%%%%%%%%%%%%%%%%%%%%%%%%%%%%%%%%%%%%%%%%%%%

%%%%%%%%%%%%%%%%% BODY OF PAPER %%%%%%%%%%%%%%%%%%

\maketitle

\section{Introduction}

Evidence for the existence of some unobservable ``dark matter'' (DM) that dominates 85\% of the Universe's matter content has been building since the early 20th Century \citep{zwicky1933redshift, rubin1978extended, rubin1980rotational}, and has now become a central pillar of the cosmological model \citep{aghanim2020planck}.
Particle physics models generally explain this mysterious matter as a massive particle that interacts only gravitationally with baryonic matter, with observations and simulations favouring a `cold and collisionless dark matter model' (hereafter referred to as CDM) \citep{copi1995big, burles1998deuterium, peacock2001measurement, clowe2006direct, aghanim2020planck, white1987galaxy,davis1982survey}.

Despite its success, CDM has faced a number of challenges in recent years.
In particular, a lack of diversity in the rotation curves of dark matter-dominated dwarf galaxies has ignited new models of non-standard DM \citep{oman2015unexpected}.
In general, astronomical DM can be modified in two ways: it can be relativistic at early times such that it free-streams out of small halos suppressing power on small scales in a top-down formation mechanism \citep{bode2001halo}, or a self-interaction in the dark sector that can create repulsive pressures, reducing large density gradients \citep{spergel2000observational}.
Both have been extensively studied in the last decade, however, in this work we will focus on Self-Interacting Dark Matter (SIDM) as a plausible model for DM.

SIDM was first proposed by \cite{carlson1992self} as a warm alternative to hot dark matter and CDM; however, \cite{de1995constraints} concluded that this version of SIDM could not match observations.
\cite{spergel2000observational} proposed an alternate version of SIDM that extends CDM through non-dissipative self-interactions to solve small-scale problems. Primarily (at the time) this aimed to solve the missing satellites and core-cusp problems.
Dark matter self-interactions are often assumed to be elastic with a mean free path of the order 1 kpc to 1 Mpc with either a velocity-independent or dependent cross-section in the range $\sigma_{\rm DM}/m=0.1-450 {\rm\ cm^2 g^{-1}}$.
Indeed it is possible to constrain SIDM in both dwarf galaxies and galaxy clusters where the mass to light ratios are extremely high.
However, in this work we will focus on the high-mass end, where SIDM can introduce perturbations to the mass distribution that can be probed by gravitational lensing. 

%Interest in SIDM faded after \cite{miralda2002test} constrained the velocity-independent cross-section down to an astronomically irrelevant $0.02 {\rm\ cm^2 g^{-1}}$ using gravitational lensing of galaxy clusters.
%However, subsequent high-resolution cosmological simulations of SIDM reassessed the constraint and found that it could be weakened to at least $0.1 {\rm\ cm^2 g^{-1}}$ \citep{peter2013cosmological}.
%Most recently, it has become apparent that the dynamics of strong SIDM in dwarf galaxies is complex, able to form cores and even `super-cusps', where the inner profile steepens beyond the characteristic NFW inner density gradient through gravo-thermal collapse \citep{gravothermal_collapse}, with observational constraints requiring $\sim100\ {\rm cm^2 g^{-1}}$ \cite{turner2021onset, yang2023strong}.
%While constraints at galaxy cluster scales remain relatively linear, requiring $\lesssim 0.3\ {\rm cm^2 g^{-1}}$ \citep{Sagunski,harvey2019observable}.
%Despite the apparent incompatibility of these limits, a velocity-{\it dependent} SIDM remains consistent with observations on all scales \citep{kaplinghat2016dark, tulin2018dark}.

Current methods for measuring the self-interaction cross-section in galaxy clusters vary depending on the dynamical state of the cluster.
Most approaches either rely on measuring the shape of the halo (since SIDM makes halos more spherical) \citep{peter2013cosmological} or the density profile of the cluster, either directly through gravitational lensing \citep{Sagunski} or indirectly through displacements between the Brightest Cluster Galaxy (BCG) and the lensing-inferred centre of the DM distribution in merging and relaxed galaxy clusters \citep{darkgiants,sirks2024,harvey2019observable}.
Despite efforts to constrain SIDM using lensing, progress is limited for three main reasons:

\begin{enumerate}
\item Observables are often degenerate with simulation nuisance parameters including feedback from the central Active Galactic Nuclei (AGN) or the softening length of the simulation \citep{wobbles_tng}.
\item Detailed strong lensing analyses of clusters that attempt to model many hundreds of multiply lensed systems are slow and often assume central density profiles derived from simulations of CDM \citep{johnson2014lens,richard2014mass}.
In the advent of stage IV telescopes such as Euclid \citep{laureijs2012euclid} and the Vera Rubin Telescope \citep{LSST}, we anticipate a significant increase in the number of galaxy cluster observations.
\item Parametric gravitational lensing models fitted to data to estimate the shape or position of halos often assume that the halos are smooth and elliptically symmetric, which can result in important information loss.
\end{enumerate}
Although there exist methods to mitigate these three hurdles, in this paper we explore the use of machine learning to efficiently extract dark matter information without the use of summary statistics in galaxy clusters.

Machine learning (ML) offers a model-agnostic approach to feature learning and parameter estimation without relying on summary statistics or specific models \citep{Hoyle2016Measuring, lanusse2023dawes}.
Neural networks (NNs) are a subtype of ML, which are composed of series of simple linear transforms combined with non-linear functions parametrised by weights optimised through an objective, the loss function, and gradient decent algorithms to create a universal function approximator, see \cite{sen2022astronomical} for a review.
The flexible nature of ML enables it to solve a broad host of problems \citep[e.g.][]{euclid_galaxy_class}, leading it to become commonplace in astronomy.
However, ML remains "black-box" in nature, with limited interpretability and transparency.
This is particularly pertinent when we are attempting to train algorithms on complex simulations and then apply to noisy data to make inferences on the nature of dark matter. 

Recently, a study applied deep learning to SIDM, training their architecture on images of galaxy clusters to classify different models of SIDM from collisionless CDM with different levels of astrophysical feedback \citep[][hereafter H24]{harvey2024darkcnn}.

They trained using the BAHAMAS-SIDM hydro-dynamic simulations \citep{mccarthy2016bahamas, robertson2019observable},
adapting the Inception-v4 convolutional NN architecture from \cite{szegedy2017inception} and \cite{merten2019dissection} achieving ${\sim}80\%$ accuracy in classifying the models CDM and SIDM with $\sigma_{\rm DM}/m=0.1$ $\rm cm^2 g^{-1}$ and $\sigma_{\rm DM}/m=1.0$ $\rm cm^2 g^{-1}$.
The performance of the NN was also tested on an unseen SIDM model with $\sigma_{\rm DM}/m=0.3$ $\rm cm^2 g^{-1}$; however, since this study used a classification architecture, the NN's output can only estimate the probabilities for the previous three models.
Therefore, to obtain an estimate for $\sigma_{\rm DM}/m$, they treated the output as a poorly sampled probability distribution function (PDF) with the expectation of the probability distribution of $n$ clusters being the estimated cross-section for the model.
The accuracy for a single galaxy cluster was $\delta\sigma_{\rm DM}/m=0.1$ $\rm cm^2 g^{-1}$ with the uncertainty decreasing by $\sqrt{n}$ for $n$ galaxy clusters.

A key limitation shared by H24 and other cosmological deep learning methods is that the hydro-simulations are fine-tuned; therefore, any algorithm trained on that data will inherently be non-general.
Hence, when applied to a new set of simulations or observations that do not reflect the simulated data, it will be forced to extrapolate, returning unreliable estimations.
In most areas of scientific research, we use empirical systematic checks to verify that the model fits the data well, for example, goodness of fit statistics.
However, ML does not have such a check.
In particular, it has no way of informing the user that the test data is related in any way to the training data, returning inferences with little insight in to how well it fits the data.

In this paper, we extend the work of H24 in two key directions.
First, we design a regression-based architecture to estimate an arbitrary cross-section along with an associated probability distribution to capture the uncertainty of the estimation.
Second, we introduce an interpretable latent space, a low-dimensional representation of the data, to address the confidently wrong problem of NNs, particularly in classification, where any input will be assigned a probability for each class without the option to say that the input is outside the training domain.
Given we utilise a variety of simulation suites, we do not follow a single cosmology, and explicitly state throughout, which one is assumed.

% The structure of the paper is as follows: in section \ref{sec:data} we describe the data we use, in section \ref{sec:method} we outline the method, including our architecture in Section~\ref{sec:architecture}, the training procedure in Section~\ref{sec:training} and our estimator in section  \ref{sec:posteriors}.
% We then present our results in section \ref{sec:results} and discuss them in section \ref{sec:dicussion} and conclude in section \ref{sec:conc}.

\section{Data}
\label{sec:data}

\begin{table*}
    \centering
    \caption{
        Table of datasets used to train and test in this work.
        From left to right we show the dataset identifier, the box length in $\rm h^{-1} Gpc$, the assumed cosmology ($^1$\cite{WMAP9cosmo}, $^2$\cite{planckParsFinal}), the DM particle mass in units of $10^9 M_\odot$, the initial gas particle mass in units of $10^9 M_\odot$, the strength of AGN feedback with respect to the fiducial, the number of clusters in the training set, the halo mass range, and the self-interaction cross-section.
        For both BAHAMAS and DARKSKIES we take four redshift slices at $z=$[0, 0.125, 0.25, 0.375].
    % and FLAMINGO we have a slightly different $z=$[0, 0.1, 0.2, 0.35, 0.45 ].
        }
    \begin{tabular}{*{9}{c}}
        \hline
        Simulation & Box Length & Cosmology & $m_{\rm dm}$ & $m_{\rm g}$ & AGN Strength & Sample \# & Mass Range & $\sigma_{\rm DM}/m$ \\
        Identifier & [{$\rm h^{-1} Gpc$}] & --- & [$\rm 10^9 M_\odot$] & [$\rm 10^9 M_\odot$] & --- & --- &[$\rm \log{M/M_\odot}$] & [$\rm cm^2 g^{-1}$] \\
        \hline
        BAHAMAS-0w & $0.4$ & WMAP9$^1$ & $5.5$ & $1.1$ & Weak & 3600 & 14.0 -- 15.5 & 0 \\
        \hline
        BAHAMAS-0 & $0.4$ & WMAP9$^1$ & $5.5$ & $1.1$ & Fiducial & 3600 & 14.0 -- 15.5 & 0 \\
        \hline
        BAHAMAS-0s & $0.4$ & WMAP9$^1$ & $5.5$ & $1.1$ & Strong & 3600 & 14.0 -- 15.5 & 0 \\
        \hline
        BAHAMAS-0.1 & $0.4$ & WMAP9$^1$ & $5.5$ & $1.1$ & Fiducial & 3600 & 14.0 -- 15.5 & 0.1 \\
        \hline
        BAHAMAS-0.3 & $0.4$ & WMAP9$^1$ & $5.5$ & $1.1$ & Fiducial & 3600 & 14.0 -- 15.5 & 0.3 \\
        \hline
        BAHAMAS-1 & $0.4$ & WMAP9$^1$ & $5.5$ & $1.1$ & Fiducial & 3600 & 14.0 -- 15.5 & 1 \\
        \hline
        \rowcolor{lightgray!50}
        DARKSKIES-0 & Zoom & Planck$^2$ & $0.068$ & $0.82$ & Fiducial & 1200 & 14.5-15.5 & 0\\
        \hline
        \rowcolor{lightgray!50}
        DARKSKIES-0.1 & Zoom & Planck$^2$ & $0.068$ & $0.82$ & Fiducial & 1200 & 14.5-15.5  & 0.1\\
        \hline
        \rowcolor{lightgray!50}
        DARKSKIES-0.2 & Zoom & Planck$^2$ & $0.068$ & $0.82$ & Fiducial & 1200 & 14.5-15.5 & 0.2\\
        \hline
    \end{tabular}
    \label{tab:datasets}
\end{table*}

A key aim of this paper is to develop an algorithm that can return a confidence measure for its estimations based on the known data in the learned latent space.
As such we require independent datasets that have differing choices of hydro-parameters, particle mass resolution (and subsequent softening), cosmologies, and box sizes.
This leads us to adopt two key suites of simulations, with an overview of the datasets shown in Table \ref{tab:datasets}.

\begin{enumerate}
    \item BAHAMAS-SIDM \citep{mccarthy2016bahamas,robertson2019observable}, as used by H24: A cosmological box with a side length of 400 Mpc ${\rm h}^{-1}$, $2\times1024^3$ particles and WMAP 9-year cosmology \citep{WMAP9cosmo}, where the particle mass is $5.5\times10^9\ {\rm M}_\odot$ for DM, and $1.1\times10^9\ {\rm M}_\odot$ for baryons.
    For redshifts ${\rm z}>3$, the Plummer-equivalent gravitational softening length is fixed in comoving coordinates; while below ${\rm z}<3$ it is $5.7$ pkpc.
    The dataset comprises three CDM models with varying levels of baryonic feedback, alongside three SIDM models with different cross-sections: BAHAMAS-0w (weak baryonic feedback), BAHAMAS-0 (fiducial CDM), BAHAMAS-0s (strong baryonic feedback), BAHAMAS-0.1 $(\sigma_{\rm DM}/m=0.1\ {\rm cm^2 g^{-1}})$, BAHAMAS-0.3 $(\sigma_{\rm DM}/m=0.3\ {\rm cm^2 g^{-1}})$, and BAHAMAS-1 $(\sigma_{\rm DM}/m=1\ {\rm cm^2 g^{-1}})$.
    Each model analyses the 300 most massive galaxy clusters across four redshift snapshots at $z=0$, $0.125$, $0.25$, and $0.375$.
    Each snapshot includes three projections, one along each principal axis, resulting in 3,600 simulated observations per simulation.
    \item DARKSKIES: A suite of SIDM zoom-in simulations \citep{harvey2025darkskies}.
    This suite of simulations mimics the initial box size of BAHAMAS by first generating an initial low-resolution DM only box of size 400 Mpc ${\rm h}^{-1}$ with $256^3$ particles and Planck cosmology \citep{planckParsFinal} to find the most massive DM halos.
    It then re-simulates a higher resolution zoom-in region with baryons.
    The zoom-in regions are cut at 5 virial radii in length.
    The simulations are super-sampled such that the DM particle mass is much lower than the gas mass: 
$m_{\rm dm}=6.8\times10^7\ {\rm M_\odot}$ and $m_{\rm g}=8.2\times10^8\ {\rm M_\odot}$ for DM and baryons respectively.
    The Plummer-equivalent comoving softening length is 2.28 pkpc.
    The dataset comprises one CDM and five SIDM models, we use the CDM and two SIDM simulations: DARKSKIES-0, DARKSKIES-0.1 $(\sigma_{\rm DM}/m=0.1\ {\rm cm^2 g^{-1}})$, and DARKSKIES-0.2 $(\sigma_{\rm DM}/m=0.2\ {\rm cm^2 g^{-1}})$.
    Each model simulates the 100 most massive galaxy clusters across four redshift snapshots at $z=0$, $0.125$, $0.25$, and $0.375$.
    Each snapshot includes three projections, one along each principal axis, resulting in a total of 1,200 simulated observations per simulation.
\end{enumerate}

In all cases the input data consists of total mass maps that include contributions from DM, stellar, gas and black hole particles, as this can be obtained from weak lensing observations of galaxy clusters and is the most useful for distinguishing between DM models.
To generate these we simply sum all the mass within each pixel to a projected depth of $10$ Mpc.
This is sufficient to account for all correlated line-of-sight structure.
We can also use the X-ray emission as an additional channel in the input to provide more information about the gas of the galaxy clusters as this will be useful in identifying differences in baryonic feedback; however, due to differences in how these maps are created between different suites of simulations, we found that this induces biases (whereas total mass maps is just the summed projected mass).
Therefore, to reduce modelling uncertainties, we focus solely on total mass maps.
Each map is binned to $\delta x = 20$ pkpc, and out to 2 Mpc, resulting in an image dimensions of $100\times100$.

Both the BAHAMAS and the DARKSKIES simulations invoke elastic, velocity-independent cross-sections that have isotropic scatterings.
To first approximation, both use a similar probabilistic scattering mechanism where they calculate the local density including the number of nearest neighbours and calculate the probability of scattering.
To ensure that the validity of the scattering equations hold, the time-steps of each simulation are reduced such that we expect there to be one scattering per time-step.
The only place the two algorithms differ is that BAHAMAS-SIDM uses a fixed radius in which to search for neighbours \citep{SIDM_bullet}, corresponding to the simulation's smoothing length, whereas DARKSKIES uses an adaptive smoothing length based on the local density, similar to smooth particle hydrodynamics \citep{correa2022}.

In Section~\ref{sec:training}, we choose to apply a logarithmic transform to $\sigma_{\rm DM}/m$; however, as CDM has a zero cross-section, we have to assign an effective $\sigma_{\rm DM}/m$.
Previous work by \cite{harvey2019observable} and \cite{roche2024brightest} found that the softening length of simulations can create a similar effect to BCG wobble as SIDM; therefore, an effective $\sigma_{\rm DM}/m$ can be calculated for CDM simulations, with $\sigma_{\rm DM}/m=0.01$ $\rm cm^2 g^{-1}$ for BAHAMAS-CDM.
While DARKSKIES-0 would have a lower effective $\sigma_{\rm DM}/m$ due to its higher resolution, we found that assigning the same classification label to all CDM simulations improves the performance; therefore, for the final results, we will use $\sigma_{\rm DM}/m=0.01\ {\rm cm^2 g^{-1}}$ for all CDM simulations.

Although the simulations are velocity-{\it independent}, H24 showed that ML algorithms trained on these are sensitive to velocity-{\it dependent} models, provided the input clusters fall within a given mass bin and the effective cross-section lies within the sampled parameter space of the trained model.

\section{Method}
\label{sec:method}

\subsection{Semi-Supervised Compact Deep Clustering}
\label{sec:cluster}

Clustering is traditionally framed as an unsupervised ML task, where the goal is to identify groups within datasets without prior knowledge of what groups the data or even the number of groups.
However, it can be extended to semi-supervised learning to inform the clustering using a few known labels.
The simplest clustering algorithm is k-means where the number of clusters, $K$, is pre-defined and cluster centroids are randomly set in the data space.
Data points are then assigned to the nearest centroid and the locations of the centroids are updated as the centre of each cluster \citep{hartigan1979algorithm}.

However, k-means and similar methods assume that meaningful clusters exist in the input space, which often fails for high-dimensional or structured data such as images.
The data must first be projected to a lower-dimensional latent space using methods such as principal component analysis \citep{pearson1901liii, hotelling1933analysis}, or neural networks \citep{ren2022deep}.
Clustering in the latent space of NNs has been successfully used to extract structure from complex data, such as in autoencoders \citep{tzoreff2018deep,yang2019deep}, or to generate pseudo labels for semi-unlabelled data \citep{kamnitsas2018semi} or completely unlabelled data \citep{caron2018deep}.

Unlike most ML problems where the target test-set may contain a variety of different labels,  for our problem, the Universe (and each simulation included in the training data) can be assumed to have only one unique class (or value of $\sigma_{\rm DM}/m$). As such, during training and testing, each simulation can be treated as a unique class with a {\it known} $\sigma_{\rm DM}/m$, with real observations of the universe treated as a single  unique class with an {\it unknown} $\sigma_{\rm DM}/m$.
In this work we use deep semi-supervised clustering methods to estimate a value for this ``universe class''.
We use the assumption that feature similarity clustering will place similar classes closer together, and dissimilar ones far apart; therefore, the label for an unknown class can be inferred by its location in the latent space.
To create an interpretable latent space capable of interpolating sparsely sampled parameter spaces, we extend the semi-supervised clustering and similarity mapping algorithm by \cite{kamnitsas2018semi}.
The algorithm takes a batch, of images, $\mat{X}\in\mathbb{R}^{N\times\ldots}$, as input and generates a set of cluster latent vectors, $\mat{Z}\in\mathbb{R}^{N\times\left|\mathcal{Z}\right|}$, and a set of class probabilities, $\mat{Y}'\in\mathbb{R}^{N\times C}$, for the set of classes, $C$, with batch size $N$ and latent space $\mathcal{Z}$.
The algorithm has three main objectives:

\begin{itemize}
    \item Map images of galaxy clusters from the same simulation to a single compact cluster in $\mathcal{Z}$.
    \item Ensure that distances between different classes in $\mathcal{Z}$ reflect the differences in cross-section between the classes.
    \item Galaxy clusters with similar physics and features should be mapped to similar locations within the latent space.
\end{itemize}

The first objective learns macroscopic features of the datasets from individual samples.
The second objective involves structuring $\mathcal{Z}$ to accurately estimate $\sigma_{\rm DM}/m$ for datasets with a known $\sigma_{\rm DM}/m$.
The third objective enables interpolation between classes, allowing us to use all datasets, including the unknown dataset, and position the datasets in $\mathcal{Z}$ based on their similarities with other datasets.

The original work by \cite{kamnitsas2018semi} designed an algorithm that has a set of data with known classes, $\left(\mat{X}_L,\mat{Y}_L\right)\sim\mathcal{D}_L$ with $\mat{X}_L\in\mathbb{R}^{N_L\times\ldots}$ and $\mat{Y}_L\in\mathbb{R}^{N_L\times C}$, and a set of data without knowledge of the classes, $\mat{X}_U\sim\mathcal{D}_U$ with $\mat{X}_U\in\mathbb{R}^{N_U\times\ldots}$.
The goal of the paper is to use a limited number of data with known classes to propagate their class to the unknown data and cluster them into a latent space for classification.
The class propagation works by first encoding the input $\mat{X}$ into a cluster latent space $\mathcal{Z}$.
A graph NN is then created from $\mathcal{Z}$ by forming connections based on the density of points with stronger connections to points close together in dense regions.
Labels can then be propagated by following the path of strongest connections which would correspond to high density areas.
The compact clustering via label propagation (CCLP) loss function is shown in Equation~\ref{eq:cclp_loss}, which is the cross-entropy between the target transition matrix, $\mat{T}\in\mathbb{R}^{N\times N}$, and $\mat{H}^{(s)}\in\mathbb{R}^{N\times N}$, where $T_{ij}$ is the transition probability between latent vectors $\bm{z}_i$ and $\bm{z}_j$ based on the propagated class labels, and $H_{ij}^{(s)}$, the probability that a Markov chain starts from $\bm{z}_i$, walks $s-1$ steps in the same class, before transitioning to $\bm{z}_j$ within $\mathcal{Z}$, see Appendix~\ref{app:cclp} and \cite{kamnitsas2018semi} for the derivation of this loss function.

\begin{figure*}
    \centering
    \includegraphics[width=\textwidth]{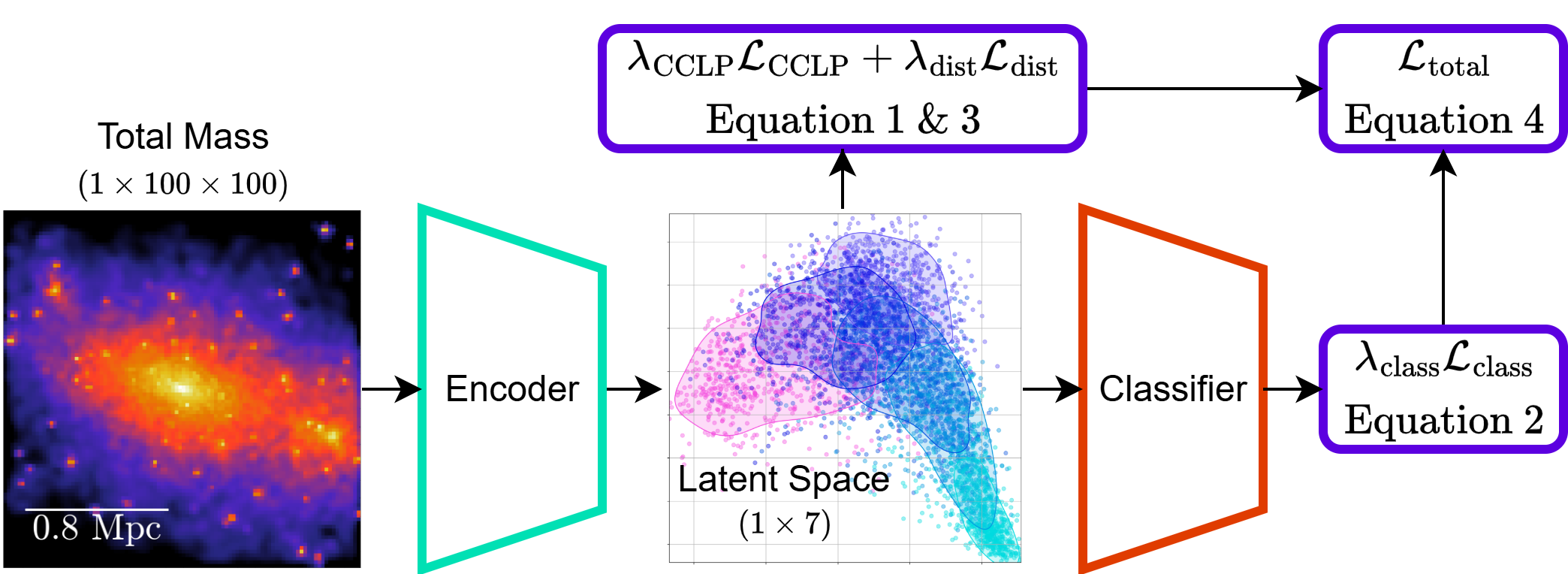}
    \caption{
    The architecture and loss functions used in this paper.
    The input is the total mass and optionally X-ray maps, which is then compressed using a convolutional NN (blue encoder) into a 7D latent space.
    From this latent space, we can get the similarity cluster and distance losses, equations~\ref{eq:cclp_loss} and~\ref{eq:dist_loss}, or further transform it using a fully connected NN (red classifier) to obtain the classification loss, Equation~\ref{eq:class_loss}.
    All losses are then weighted summed using the weights $\lambda_{\rm CCLP}$, $\lambda_{\rm dist}$, and $\lambda_{\rm class}$, Equation~\ref{eq:total-loss}
    See Figure~\ref{fig:network-architecture} for the full encoder and classifier architecture.
    }
    \label{fig:architecture}
\end{figure*}

\begin{equation}
    \label{eq:cclp_loss}
    \mathcal{L}_{\rm CCLP} = \frac{1}{SN^2}\sum_{s=1}^S\sum_{i=1}^N\sum_{j=1}^N-T_{ij}\log{H_{ij}^{(s)}}
\end{equation}

Finally, a classifier can be trained using the classification loss, shown in Equation~\ref{eq:class_loss}, with these propagated classes.
This loss function is a cross-entropy loss which takes class probability predictions, $\mat{Y}'$, and the one-hot encoding of the ground truth, $\mat{Y}\in\mathbb{R}^{N\times C}$.
The classification loss supports the creation of efficient clusters and implicit learning of differences between classes not provided to the NN, such as baryonic feedback.
Classification can be used for all classes, including classes with the unknown labels, as we know each simulation or observation an image comes from; therefore, all observations from a single simulation or observation would have the same cosmological parameters.

\begin{equation}
    \label{eq:class_loss}
    \mathcal{L}_{\rm class} = -N^{-1}\sum_{i=1}^N\sum_{c\in C_l} Y_{ic}\log{Y'_{ic}}
\end{equation}

While we do not have any unknown classes due to the nature of our problem, we can take advantage of the clustering regulariser, Equation~\ref{eq:cclp_loss}, as \cite{kamnitsas2018semi} found that even in the case where all labels were known, their algorithm still improved over traditional classification algorithms and provides a way to enforce structure within the latent space.

In addition to these two loss functions, we also add a mean squared error (MSE) loss, shown in Equation~\ref{eq:dist_loss}, where $\mat{M}_c\in\mathbb{R}^{\left|\mathcal{Z}\right|}$ is the centre of the cluster for class $c\in C_l$ and $\sigma_c$ is $\sigma_{\rm DM}/m$ for that class, which ensures that the first dimension of $\mathcal{Z}$ is physicalised and corresponds to $\sigma_{\rm DM}/m$.

\begin{equation}
    \label{eq:dist_loss}
    \mathcal{L}_{\rm dist} = \left|C_l\right|^{-1}\sum_{c\in C_l}\left(M_{c,1} - \sigma_c\right)^2
\end{equation}

We then combine the three loss functions using a weighted sum to form the total loss, Equation~\ref{eq:total-loss}.
The weights used are: $\lambda_{CCLP}=2.2$, $\lambda_{dist}=1$, and $\lambda_{class}=1$ for each corresponding loss, the weights chosen were found to give the best $\sigma_{\rm DM}/m$ accuracy for unknown validation datasets.
Our choices of hyper-parameters and training decisions are discussed in appendix~\ref{sec:biases}.
% Therefore, the total loss is the weighted sum of the three loss functions weighted by three hyperparameters $\lambda_{CCLP}=2.2$, $\lambda_{dist}=1$, and $\lambda_{class}=1$ for each loss respectively, Equation~\ref{eq:total-loss}, where the values chosen were found to give the best $\sigma_{\rm DM}/m$ accuracy for unknown datasets.

\begin{equation}
    \label{eq:total-loss}
    \mathcal{L}_{\rm total} = \lambda_{\rm CCLP} \mathcal{L}_{\rm CCLP}+\lambda_{\rm class}\mathcal{L}_{\rm class}+\lambda_{\rm dist}\mathcal{L}_{\rm dist}
\end{equation}

\subsection{Architecture}
\label{sec:architecture}

Figure~\ref{fig:architecture} shows a summary of the architecture used in this work, for a full description see Appendix \ref{app:architecture}
The inputs are total mass maps, and optionally X-ray maps, as described in Section~\ref{sec:data}.
Each input map belongs to a specific dataset, such as a simulation, which is associated with a classification label and, if known, a value for $\sigma_{\rm DM}/m$.
The inputs belonging to the unknown dataset that we want to find $\sigma_{\rm DM}/m$ for can also be passed through the NN during training along with a unique label; however, evidently the value of $\sigma_{\rm DM}/m$ would be unknown for this class.
The inputs are then passed through a convolutional NN, the encoder (turquoise, trapezium), that learns the features from the input and produces a $\left|\mathcal{Z}\right|=7$ dimensional latent space.
The aim is to make this latent space learn the most important features in the data, primarily $\sigma_{\rm DM}/m$, and any other differences between datasets.
In the latent space, each dataset forms a distinct cluster, with individual inputs from the same dataset grouped closely together.
We show in Figure~\ref{fig:architecture} the 68\% contours for visualisation for each dataset or class represented by a different colour.
The location of each cluster in the latent space will correspond to the physical properties of the dataset relative to the other datasets, enforced by the classifier (red, trapezium) and three loss functions described in Section~\ref{sec:cluster}.

% \begin{itemize}
%     \item $\mathcal{L}_{\rm dist}$ enforces that the first dimension of the latent space is equal to $\sigma_{\rm DM}/m$ for all datasets where this is known.
%     \item $\mathcal{L}_{\rm CCLP}$ enforces that inputs that belong to the same dataset should be close together in the latent space, while different datasets should be far apart.
%     \item The final loss function, $\mathcal{L}_{\rm class}$, is calculated from the outputs of an additional NN, the classifier, that learns the probabilities that the input belongs to each dataset used during training, which helps to enforce compact clusters in the latent space for each dataset, minimising confusion between datasets.
% \end{itemize}

% The final loss is the weighted sum of these three loss functions with weights: $\lambda_{\rm dist}$, $\lambda_{\rm CCLP}$, and $\lambda_{\rm class}$, respectively.

The encoder does the bulk of the computation, and therefore, has the most complex architecture.
The architecture follows the standard shape found in many convolutional NNs where the number of convolutional filters in a layer increase as the input is downscaled; however, the main difference is that we stack blocks of Inception modules \citep{szegedy2015going} rather than individual convolutional layers and use max pooling for downscaling the input.
After the Inception and max pool layers, there are two linear layers to form the cluster latent space.
The classifier is much simpler and is composed of just two linear layers.
To prioritise the most important features in the first dimensions, we add an information ordered bottleneck layer \citep{ho2023information} between the encoder and latent space.
The information ordered bottleneck zeros out all latent dimensions greater than $k$, where $k$ is a random integer given by $1\leq k<\left|\mathcal{Z}\right|$, this results in the first dimensions representing the most important features as these will have the greatest chance of contributing to the loss function without being zeroed-out, with subsequent dimensions representing less important features.
To introduce non-linearity into the network, we use the following activation functions: exponential linear units (ELUs) \citep{clevert2015fast} for convolutional layers, and scaled exponential linear units (SELUs) \citep{klambauer2017self} for linear layers.
For convolutional layers, we use a dropout probability of 5\% to reduce overfitting \citep{srivastava2014dropout}.

To save on computational resources, some of our tests use a smaller encoder architecture that only uses standard convolutional layers instead of Inception modules, enabling us to perform more tests and increase ensemble learning to reduce NN stochastic uncertainties.
We use the same activation functions as for the full architecture, but we change the dropout probability to 10\% for both convolutional and linear layers.
This reduced NN will be used in the tests performed during the method section of this paper, while the full NN will be used for the results section, unless otherwise stated.
The total number of parameters for the full architecture is 3,443,156, while the number of parameters in the reduced architecture is 232,204.

\begin{figure}
    \centering
    \includegraphics[width=\columnwidth]{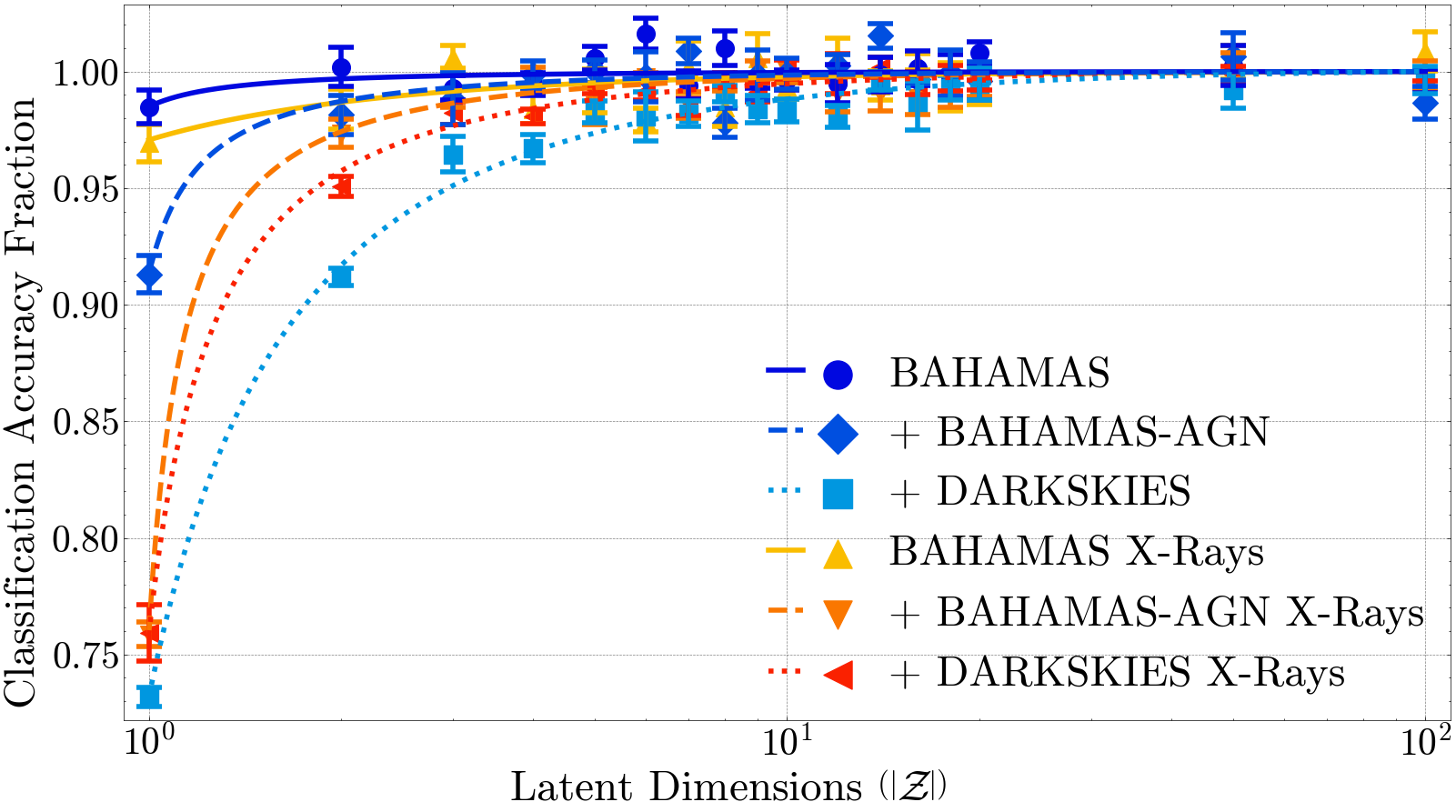}
    \caption{
        Average classification accuracy from five NNs, normalised to the asymptotic accuracy, against the number of latent dimensions.
        Two sets of NNs are trained, one with X-rays included (reds/triangles) and the other excluded (blues/non-triangles).
        Each line shows an increase in the number of simulations included in the training, starting with BAHAMAS-0 and BAHAMAS-SIDM (solid), then BAHAMAS-AGN (dashed), and finally DARKSKIES (dotted).
        We fit each set of data with an $y=a+\arctan\left( \left(\left|\mathcal{Z}\right|-b\right)/c\right)$ fit (lines).
    }
    \label{fig:latent-dimensions}
\end{figure}

We chose $\left|\mathcal{Z}\right|=7$ to balance interpretability (which favours lower dimensionality) and network accuracy (which benefits from higher dimensionality).
To determine the optimal number of dimensions, we trained five reduced NNs per combination of latent dimensions, X-ray inclusion, and number of simulations.
We train NNs with datasets starting with just BAHAMAS-0 and BAHAMAS-SIDM (solid), then adding BAHAMAS-AGN (dashed), and finally adding DARKSKIES (dotted).
Figure~\ref{fig:latent-dimensions} shows the average classification accuracy, normalised to the asymptotic accuracy, against the number of latent dimensions for the two datasets. 
To determine the point of diminishing returns we fit an arctan function, $y=a+\arctan\left( \left(\left|\mathcal{Z}\right|-b\right)/c\right)$ to each training set.
% We find before normalisation that $a=-1.114\pm0.001$, $b=0.967\pm0.066$, $c=(1.53\pm3.03)\cdot10^{-3}$ for BAHAMAS, $a=-1.028\pm0.001$, $b=0.754\pm0.054$, $c=(4.03\pm0.89)\cdot10^{-2}$ for BAHAMAS+AGN and $a=-1.053\pm0.001$, $b=0.710\pm0.038$, $c=(5.48\pm0.66)\cdot10^{-2}$ for BAHAMAS+AGN+DARKSKIES.
We find that the number of the latent dimensions, $\left|\mathcal{Z}\right|$, required to achieve 98\% of the asymptotic accuracy for BAHAMAS without X-rays is 0.9, if we add X-rays, this then increases to 1.4.
If we add BAHAMAS AGN, we find $\left|\mathcal{Z}\right|$ increases to 1.6 and 2.3 for X-rays excluded and included, respectively.
Finally, adding DARKSKIES further increases $\left|\mathcal{Z}\right|$ to 6.3 and 3.3 for X-rays excluded and included, respectively.
These results demonstrate that the required dimensionality grows with the degrees of freedom.
The main outlier lies with the addition of DARKSKIES where excluding X-rays requires a larger $\left|\mathcal{Z}\right|$, this is likely due to the difference in X-ray maps between BAHAMAS and DARKSKIES reducing the maximum classification accuracy from 59\% to 48\% with the exclusion of X-rays; therefore, flattening the best fit and increasing $\left|\mathcal{Z}\right|$ required to achieve 98\% of asymptotic accuracy.
Since $\left|\mathcal{Z}\right|\sim7$ achieves at least 98\% of the asymptotic accuracy for all scenarios, we adopt this value for the remainder of the paper.

Previous work by H24 used Inception-v4; however, the motivation for moving away from Inception-v4 in this work to a custom architecture is that its performance in regression-based tasks for our data was sub-optimal, likely due to the large downscaling effect of the stem module that takes the input images of dimension $\left(100\times100\right)$ down to $\left(10\times 10\right)$; therefore, almost all spatial information is lost at the beginning of the NN, leading to our more controlled downscaling, allowing more spatial information to be captured.
However, more advanced architectures could be explored for potentially improved performance such as ConvNeXt \citep{liu2022convnet} and vision transformers \citep{liu2021swin}.

\subsubsection{Training}
\label{sec:training}

We split the data into training and validation sets in a 4:1 ratio to prevent overfitting.
The NN is trained on the training set by minimising the loss and updating its parameters via backpropagation, while the validation set is used to evaluate generalisation performance of the NN on unseen data and tune hyperparameters.
We treat the validation fraction of the unknown dataset as the test dataset that we want to obtain $\sigma_{\rm DM}/m$ for, as our method can leverage information from the unknown dataset to help inform its estimations.
We exclude the DARKSKIES simulations from hyperparameter tuning to prevent any biases in our results.

We use the \texttt{PyTorch} \citep{Ansel_PyTorch_2_Faster_2024} library for implementing NNs, with the AdamW optimiser \citep{loshchilov2017fixing} for stochastic gradient descent with a learning rate of $10^{-4}$.
We implemented a learning rate scheduler that reduces the learning rate by half if the validation loss does not improve by at least 10\% over ten epochs, which encourages fast initial convergence, while enabling more precise optimisation near minima.
In order to improve convergence time, we normalise the data to be of order one, with the input image maps each normalised to a maximum value of one and for $\sigma_{\rm DM}/m$, we take the logarithmic transform and normalise so that $\log{\left(\sigma_{\rm DM}/m\right)}$ lies between zero and one.
We train with a batch size of 120 on an NVIDIA RTX 4080 GPU with 16 GB of video memory.
Due to the reduced number of parameters in the reduced NN, the computational time is reduced by a factor of three, enabling us to carry out more calibration tests.

\subsection{Estimating $\sigma_{\rm DM}/m$ and Confidence}
\label{sec:posteriors}

\begin{figure}
    \centering
    \includegraphics[width=\columnwidth]{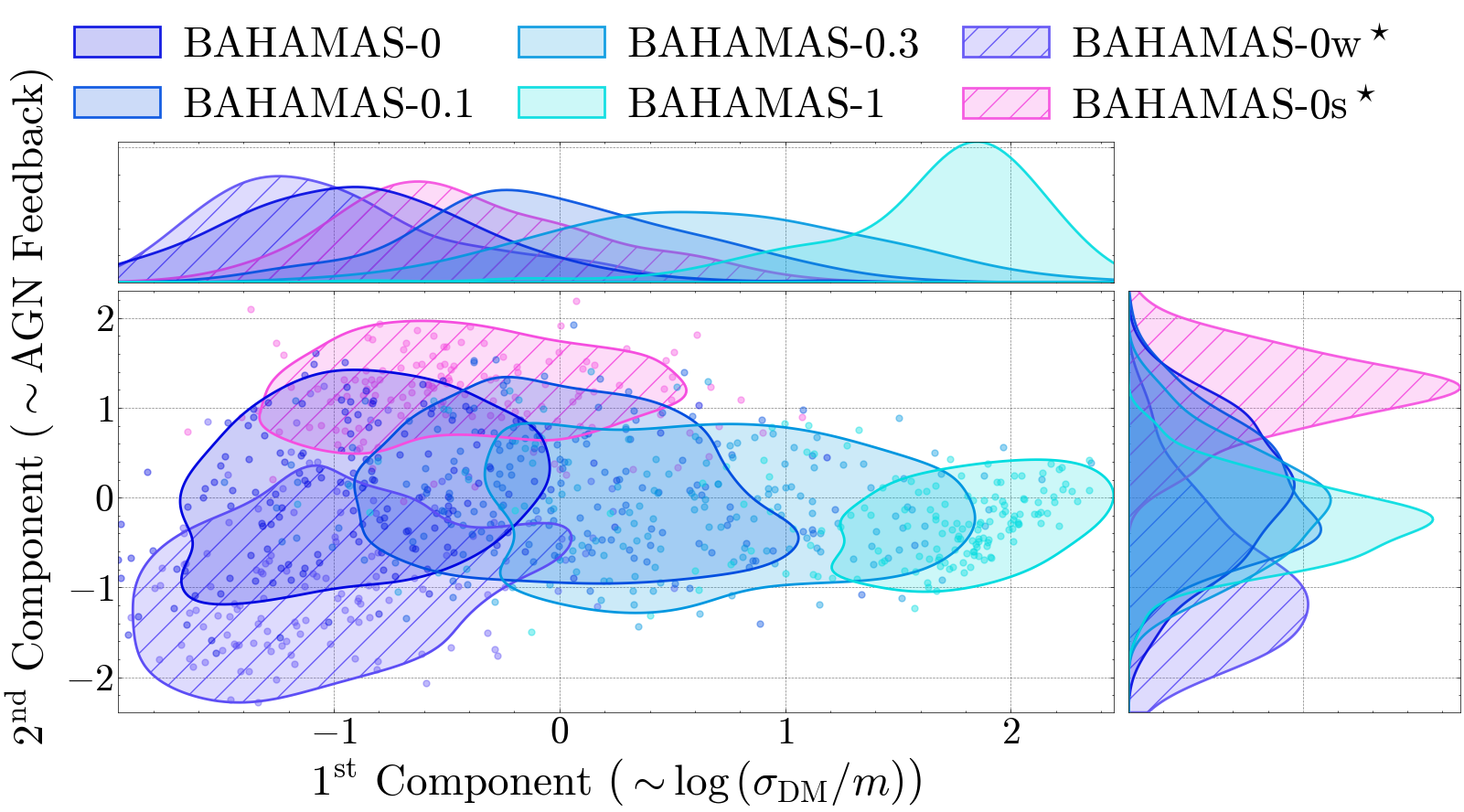}
    \caption{
    The first two components of the PCA of the 7D latent space.
    Each point corresponds to a galaxy cluster from its colour corresponding simulation with the contours representing the 68\% region for that simulation.
    The unknown datasets, BAHAMAS-0w and BAHAMAS-0s, are represented by the hatched contours and asterisk next to the legend label.
    We interpret that the first component corresponds to a transformation of $\log{\left(\sigma_{\rm DM}/m\right)}$ and the second component corresponds to the level of AGN feedback.
    }
    \label{fig:bahamas-agn-pca}
\end{figure}

\begin{figure*}
    \centering
    \includegraphics[width=0.49\textwidth]{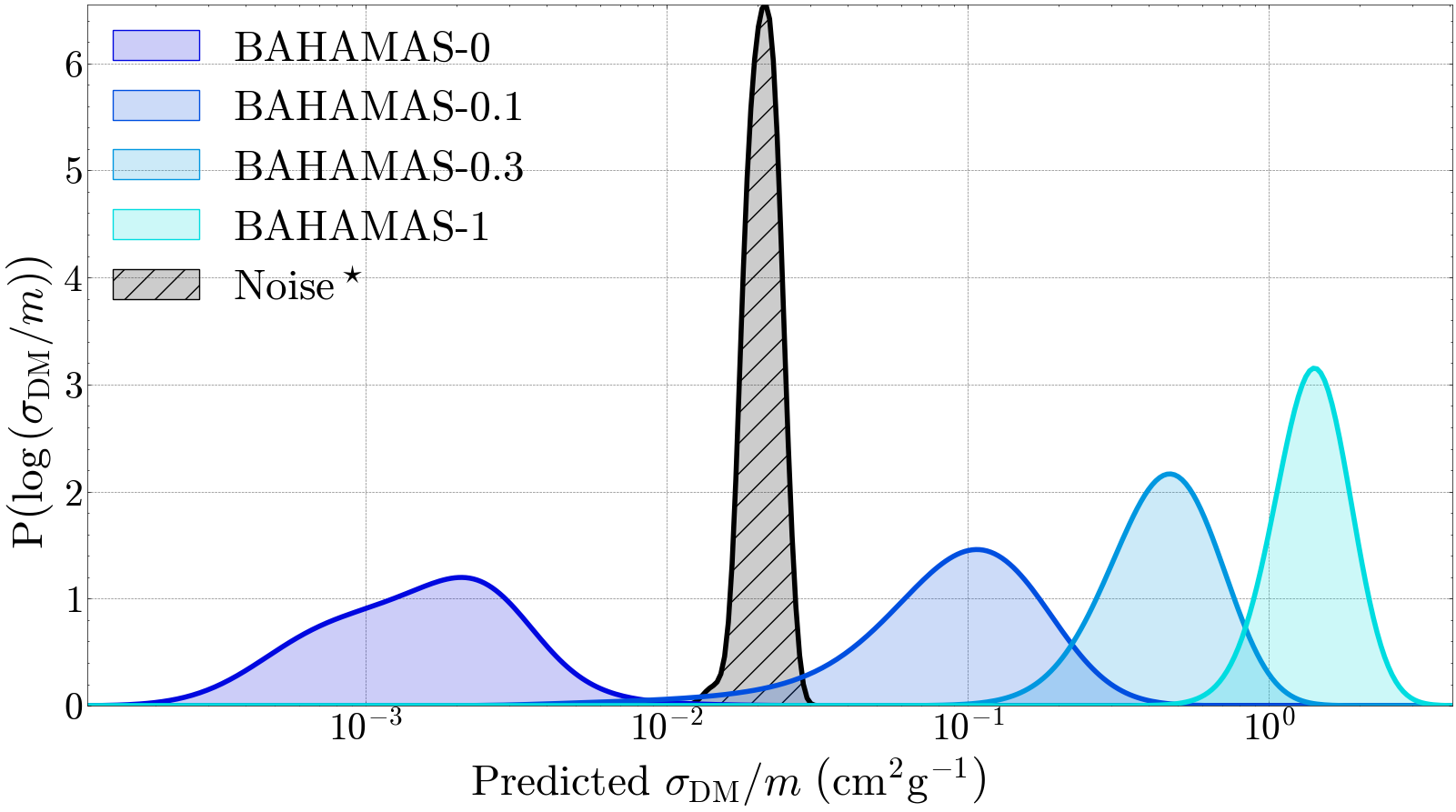}
     \includegraphics[width=0.49\textwidth]{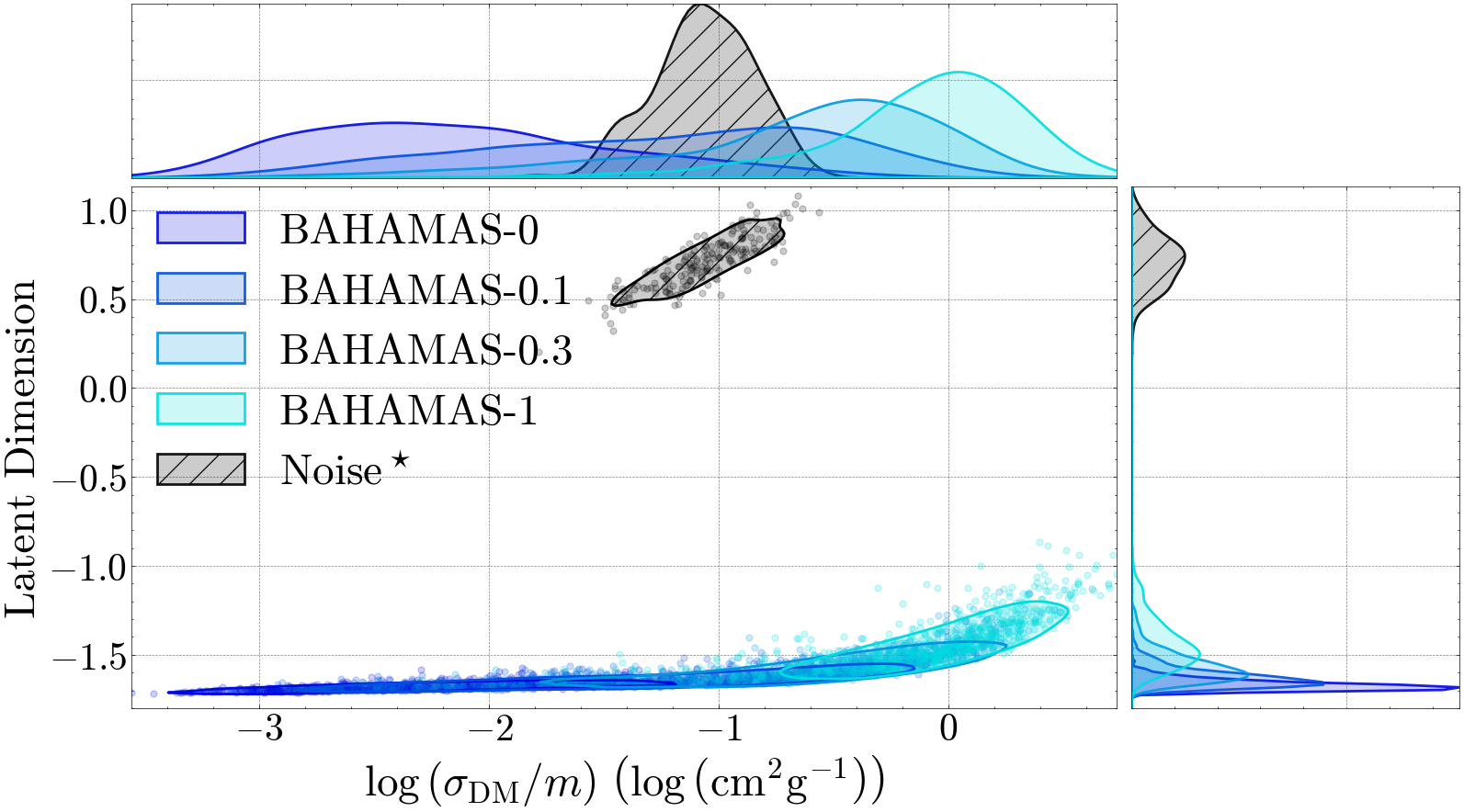}
    \caption{
        Method for building confidence in our ML estimator for data outside the training domain.
        {\it Left:} We train an ensemble of 10 regression NNs on the fiducial hydro BAHAMAS simulations.
        We show the combined probability distribution functions (PDF) of $\log{\left(\sigma_{\rm DM}/m\right)}$ for the known BAHAMAS simulations (blue shades) and blind uniform random noise (black hatched).
        We find that the regressor consistently estimates a significant, positive cross-section of $\sim0.6\ {\rm cm^2 g^{-1}}$ with no regard to its confidence, presenting the issue with direct regression estimators.
        {\it Right:} We show the first and third latent dimensions from our compact clustering algorithm trained with the same fiducial BAHAMAS as known (blue shades) and the random noise dataset as unknown (black hatched).
        The first latent dimension corresponds to $\log{\sigma_{\rm DM}/m}$, which we would naively assume that the noise dataset has a cross-section of $\sim0.1\ {\rm cm^2g^{-1}}$; however, from the third dimension we see that the noise dataset shares no similarities with the known simulations and therefore, cannot be trusted.
    }
    \label{fig:noise-regression}
\end{figure*}

For each galaxy cluster in the validation set we get an estimate for $\sigma_{\rm DM}/m$, with all galaxy cluster estimations in one simulation forming a distribution for $\sigma_{\rm DM}/m$, representing uncertainty due to data variation.
We then calculate the PDF using a Gaussian kernel.
However, NNs introduce additional uncertainty due to their stochastic nature, including randomly weight initialisation, dropout, and mini-batches.
To capture the NN uncertainty, we employ ensemble learning \citep{dasarathy2005composite, dong2020survey}, training multiple NNs with different initialisations.
We then combine the NNs estimated PDFs for $\sigma_{\rm DM}/m$ by taking the product to account for NN uncertainty.

In order to quantify our confidence in an estimation, we can use the more expressive latent space to identify whether a test dataset falls outside the training domain, resulting in untrustworthy estimations.
Qualitatively, if a test dataset forms a cluster far from those of the training datasets in the latent space, we can conclude that there are few if any shared features, suggesting extrapolation was performed and the dataset lies outside the training domain.
While the information ordered bottleneck orders the latent space in terms of importance, there are still degenerate dimensions; therefore, we can also apply principal component analysis (PCA) \citep{pearson1901liii, hotelling1933analysis} to more effectively visualise the dominant features.
PCA works by finding a new set of orthogonal bases, where the first basis explains the most variance in the data, followed by subsequent dimensions.

To quantify the confidence that the test set lies within the training domain, we project all non-physical latent dimensions (i.e. all but the first), along the axis connecting the centroid of the test set to each known dataset.
We exclude the first dimension to avoid biasing the estimate of $\sigma_{\rm DM}/m$ towards any particular simulation.
We assess the confidence by measuring the degree of overlap between the known and test projected distributions.
We also experimented with different confidence metrics, including the Earth mover distance \citep{rubner1997earth, rubner2000earth}, KS-test, and NN classification accuracy; however, the overlap proved the most interpretable and gave the largest dynamic range.

\section{Results}
\label{sec:results}

In this section we present the results from variety of tests. The aim of this section is to present how compact clustering can lead to robust and confident estimates of the cross-section, moreover it can aid interpretability.

\subsection{Self-Organising Latent Space for Robust Inference}

\begin{figure*}
    \centering
    \includegraphics[width=\columnwidth]{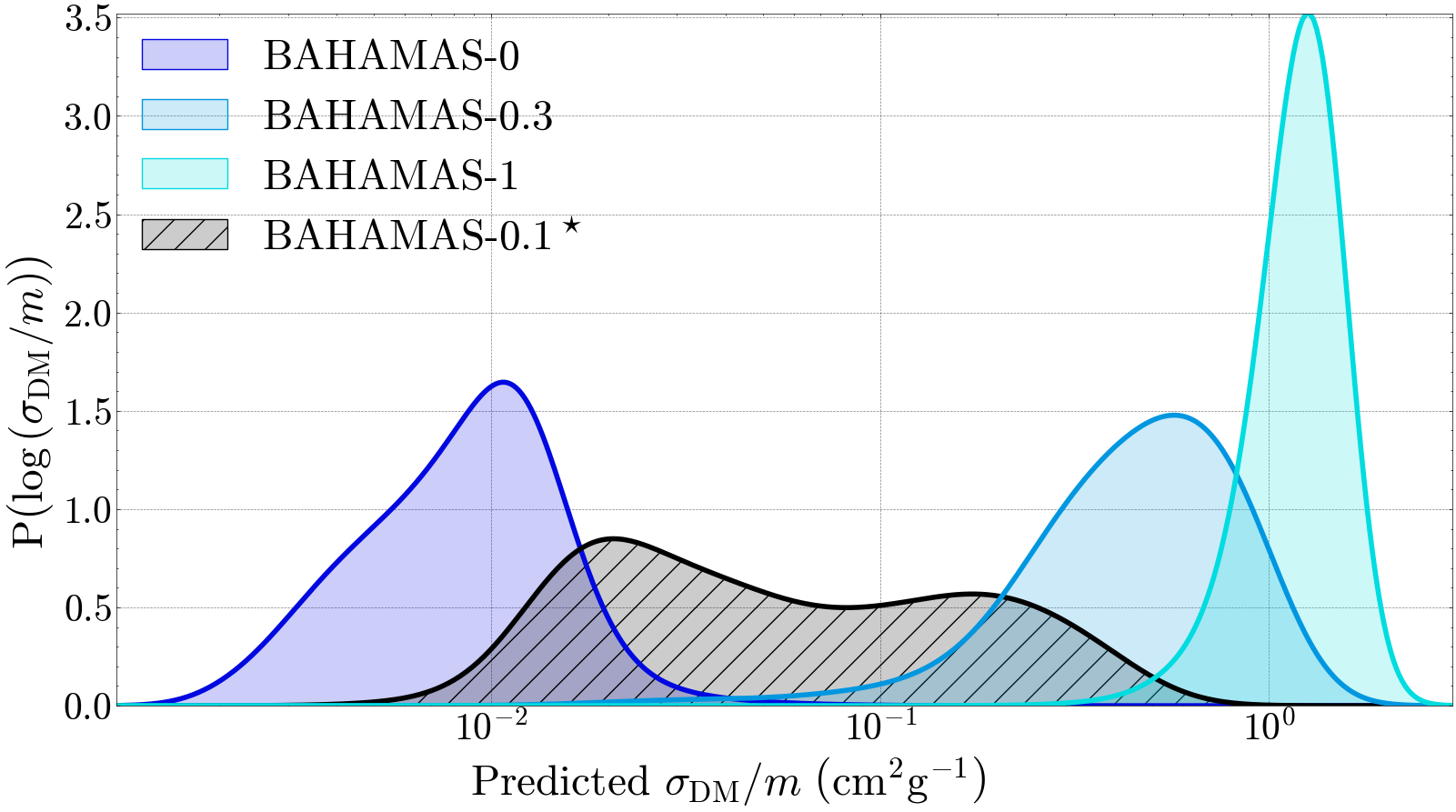}
        \includegraphics[width=\columnwidth]{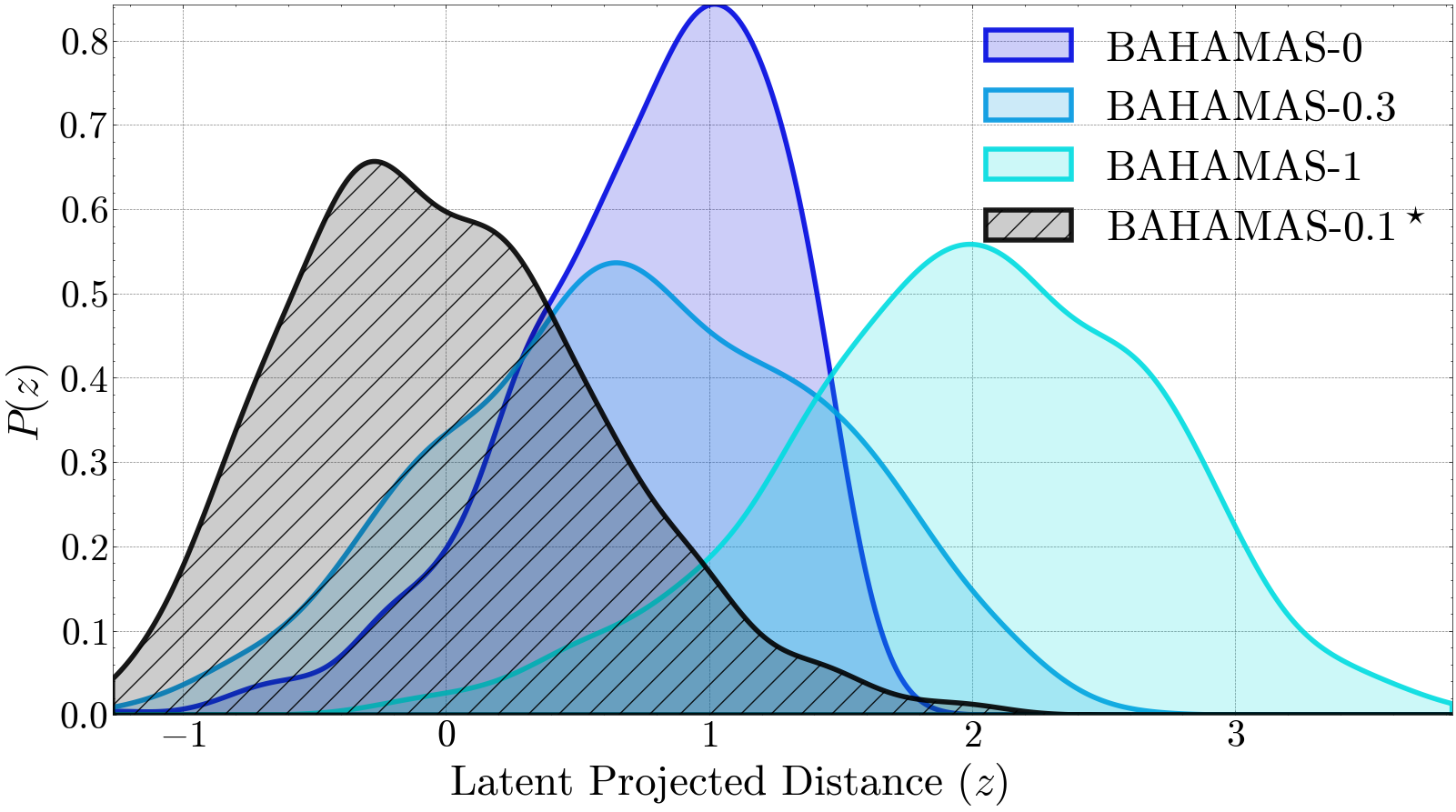}
    \caption{
        A consistency check on in-distribution testing.
        {\it Left:} We train an ensemble of 3 clustering NNs and show the combined PDF of $\log{\left(\sigma_{\rm DM}/m\right)}$ for the known BAHAMAS-0, BAHAMAS-0.3, and BAHAMAS-1 (solid blue shades) and unknown BAHAMAS-0.1 (black dashed).
        We find a cross-section of $\sim0.05\ {\rm cm^2 g^{-1}}$ and within $1\sigma$ of $0.1\ {\rm cm^2 g^{-1}}$ for BAHAMAS-0.1.
        {\it Right:} The projected 7D latent space from our clustering algorithm into a 1D distance PDF, where each known simulation (blue shades) is projected in the direction from the centre of their distribution to the centre of BAHAMAS-0.1.
        BAHAMAS-0.1 (black hatched) is projected in the direction of BAHAMAS-0.3 to show the greatest overlap.
        The distance has arbitrary units as it depends on the scale of the latent space.
        BAHAMAS-0.1 shares a large overlap of 49\% and 55\% with BAHAMAS-0 and BAHAMAS-0.3, respectively, leading to the conclusion that it lies within the training domain.
    }
    \label{fig:bahamas-0.1-distribution}
\end{figure*}

To demonstrate the interpretability of our model, we first train and test on BAHAMAS only.
In this case (and only this case), we use two input channels, the total mass maps, but also an additional X-ray channel.
The X-ray maps are taken from H24, which used methods from \cite{BAHAMASxray} to generate realistic X-ray emission maps.
We train on all fiducial BAHAMAS models (BAHAMAS-0 and BAHAMAS-SIDM), and we treat both the strong and weak AGN variants (BAHAMAS-0s and BAHAMAS-0w respectively) as unknown.

Figure \ref{fig:bahamas-agn-pca} shows the first two components from PCA.
The training data are shown with solid contours and projected histograms, while the unlabelled test data use hatched contours and histograms.
We also denote the test data with an asterisks in the legend.
Due to $\mathcal{L}_{\rm dist}$, the optimisation process organises the latent space such that training samples with different $\sigma_{\rm DM}/m$ extend along the first principal component.
The estimated $\sigma_{\rm DM}/m$ for BAHAMAS-0w and BAHAMAS-0s is $\sigma_{\rm DM}/m=1.46^{+14.95}_{-1.23}\times10^{-3}\ {\rm cm^2g^{-1}}$, $\sigma_{\rm DM}/m=6.86^{+25.15}_{-5.03}\times10^{-2}\ {\rm cm^2g^{-1}}$ respectively, with BAHAMAS-0 having $\sigma_{\rm DM}/m=1.06^{+4.45}_{-0.91}\times10^{-2}\ {\rm cm^2g^{-1}}$, resulting in over an order of magnitude difference in estimated $\sigma_{\rm DM}/m$ for the three CDM simulations.
The unlabelled test data with zero cross-section aligns with the BAHAMAS-0 cluster in the first component; however, subtle differences cause the strong and weak AGN variants to deviate from BAHAMAS-0, extending primarily along the second principal component, orthogonal to $\sigma_{\rm DM}/m$.
Looking at the second component, we see BAHAMAS-SIDM and BAHAMAS-0 line along the same line with BAHAMAS-0 and BAHAMAS-1 having values of $0.07^{+0.81}_{-0.85}$ and $-0.29\pm0.45$ respectively, while BAHAMAS-0w and BAHAMAS-0s have offsets of $-1.01^{+0.95}_{-0.68}$ and $1.16^{+0.41}_{-0.50}$ respectively.
The first two components have an explained variance ratio of 47\% and 28\% respectively, for a total explained variance ratio of 75\%.

This reveals two key insights, first, the second principal component captures the level of AGN feedback in the simulations without any prior labelling.
Second, the extracted features appear to be orthogonal to those associated with $\sigma_{\rm DM}/m$, similar to what H24 found.
Since our primary focus is on estimating confidence in $\sigma_{\rm DM}/m$, we omit the X-ray channel moving forward, as it only contributes to interpretability and not constraining power in $\sigma_{\rm DM}/m$.
Future work may explore the use of X-rays to study AGN feedback in clusters; however, this is beyond the scope of this work.

\subsection{Testing on Out-of-Domain Data: Random Noise}

Having demonstrated the flexibility of our self-organising latent space, we now test its ability to recognise out-of-domain (OOD) features.
In traditional regression models (c.f H24), estimations are often accurate and precise on training and in-domain test data; however, if the test set is OOD, these models lack any mechanism to indicate this mismatch, resulting in estimations that cannot be trusted.
As a result, they often return overly confident estimations.

To test this hypothesis, we construct a simple test case using completely random noise, where each pixel is sampled from a uniform distribution between 0 and 1, and hence, contains no signal.
We expect that this will manifest as anomalous structure in the latent space, whereas traditional regression would return an estimate regardless.
We train both the clustering-based and traditional regression models using a reduced architecture, training each model ten times for 150 epochs.
The architectures are identical up to the final layer and the information ordered bottleneck.
In the clustering-based model, the final layer outputs a probability for each class, with $\sigma_{\rm DM}/m$ derived from the latent space.
In the regression model, the final layer directly outputs a single $\sigma_{\rm DM}/m$ value.
The NNs are trained on the fiducial BAHAMAS simulations only.
For the clustering regression the noise is treated as unknown during training, while it is not used during training in the traditional regression NN.

The left panel of Figure~\ref{fig:noise-regression} shows the combined estimated distributions from the ten runs from traditional regression, with the BAHAMAS simulations shown in shades of blue and noise shown in hatched black.
The NN estimates $0.021^{+0.004}_{-0.003}\ {\rm cm^2 g^{-1}}$, $0.052^{+0.022}_{-0.014}\ {\rm cm^2 g^{-1}}$, $0.350\pm0.070\ {\rm cm^2 g^{-1}}$, and $0.612^{+0.082}_{-0.080}$ for BAHAMAS-0, BAHAMAS-0.1, BAHAMAS-0.3, and BAHAMAS-1 respectively.
For the noise dataset, it is incorrectly assigned a physically meaningful value of $0.595^{+0.052}_{-0.048}\ {\rm cm^2 g^{-1}}$.

In contrast, the clustering model allows us to examine the latent space directly.
The right panel of Figure~\ref{fig:noise-regression} shows the $\sigma_{\rm DM}/m$ dimension and the third latent dimension for different datasets, for one representative run.
We choose the third latent dimension as this shows the greatest separation between the noise and the BAHAMAS simulations.
From the top left subplot, which corresponds to $\sigma_{\rm DM}/m$, we find $\sigma_{\rm DM}=0.118\pm0.007\ {\rm cm^2g^{-1}}$ for the noisy dataset taken from ten runs.
The third dimension, far right subplot, shows clear separation between the noise and BAHAMAS simulations with no shared features, confirming that the noise lies outside the training domain.
We quantitively measure this by projecting the distribution of values in the direction from the centre of each BAHAMAS cluster to the centre of the noisy data.
%Figure~\ref{fig:noise-mmd} more clearly shows the separation between the known BAHAMAS simulations and the unknown noise dataset,
%Then, we calculate the average EMD distance of $3.76\pm0.07$ for the four known BAHAMAS simulations, with a minimum of $3.64$ for BAHAMAS-0.1, while if we measure the EMD for BAHAMAS-0 compared to the other BAHAMAS simulations, we get $1.65\pm0.45$, with a minimum of $0.90$ for BAHAMAS-0.1 and a maximum of $2.46$ for BAHAMAS-1.
This projection allows us to estimate the overlap between the noise and each simulation cluster. 
We find that the overlap is consistent with zero, confirming that the noise dataset is recognised as OOD.

%\begin{figure}
%    \centering

%     \caption{
%     We show the projected 7D latent space from our clustering algorithm into a 1D distance PDF, where each known simulation is projected in the direction from the centre of their distribution to the centre of the noise distribution (blue shades).
%     The noise distribution is projected in the direction of BAHAMAS-0 (black hatched).
%     The distance has arbitrary units as it depends on the scale of the latent space.
%     The noise dataset shows no overlap with the known simulations, leading to the conclusion that the noise is foreign to the NN.
%     }
%     \label{fig:noise-mmd}
% \end{figure}

\subsection{Confidence on In-Domain Data}

Having shown that our compact clustering method can self-organise its latent space to produce a confidence estimate in our estimations, we move to a more realistic scenario.
To evaluate the model's ability to report confidence in realistic OOD test data, we first assess its performance on an in-domain test set.
Specifically we train on the fiducial BAHAMAS simulations with BAHAMAS-0.1 treated as unknown.
We train the full architecture three times for 150 epochs each to construct an ensemble.
The left hand panel of Figure \ref{fig:bahamas-0.1-distribution} shows the results of the validation set on the known simulations (solid colours) and the unknown test set, BAHAMAS-0.1 (black hatched).
The legend marks the test label with an asterisk.
The NN estimates $\sigma_{\rm DM}/m$ of $(8.63^{+5.92}_{-4.33})\cdot10^{-3}\ {\rm cm^2g^{-1}}$, $0.474^{+0.354}_{-0.238}\ {\rm cm^2g^{-1}}$, and $1.19^{+0.33}_{-0.32}\ {\rm cm^2g^{-1}}$ for BAHAMAS-0, BAHAMAS-0.3, and BAHAMAS-1 respectively, while for the unknown BAHAMAS-0.1, the NN estimates $0.048^{+0.158}_{-0.031}\ {\rm cm^2g^{-1}}$.
However, the BAHAMAS-0.1 is bimodal, which leads to the wide confidence range.
Due to our feature-based clustering, several physical features correlate with the effective strength of self-interaction, such as the mass of the galaxy cluster or the dynamical state.
This results in some SIDM0.1 halos (low mass halos for example) appearing indistinguishable from CDM.
Since the loss function acts to separate clusters based on feature similarity, it can  create two populations resulting in a bimodal distribution for SIDM0.1.
This is more pronounced when the unphysical choice for CDM is very far from 0.1, thus dragging those low-mass halos in SIDM0.1 away from its true value and creating a bimodal distribution.
The right hand panel of Figure \ref{fig:bahamas-0.1-distribution} shows the one-dimensional projected latent space.
We see that the in-domain unknown test set has a significant overlap of $48.8\pm0.3\ \%$ and $55.3\pm1.0\ \%$ with the BAHAMAS-0 and the BAHAMAS-0.3 simulations, respectively, reflecting the model's confidence in its estimation and its similarity to these datasets.

We repeat the process with BAHAMAS-0.3 treated as the unknown dataset and BAHAMAS-0.1 included in the training set.
The NN estimates $\sigma_{\rm DM}/m$ of $(5.86^{+8.28}_{-3.01})\cdot10^{-3}\ {\rm cm^2g^{-1}}$, $0.112^{+0.205}_{-0.082}\ {\rm cm^2g^{-1}}$, and $1.18^{+0.36}_{-0.31}\ {\rm cm^2g^{-1}}$ for BAHAMAS-0, BAHAMAS-0.1, and BAHAMAS-1 respectively, while for the unknown BAHAMAS-0.3, the NN estimates $0.650^{+0.505}_{-0.381}\ {\rm cm^2g^{-1}}$.
We see that the unknown BAHAMAS-0.3 also has a significant overlap of $59.9\pm0.3\ \%$ with the BAHAMAS-0.1 simulation, higher than for BAHAMAS-0.1 as the unknown, possibly explaining the lower log uncertainties and non-bimodal posterior.
% The minimum EMD is $0.188\pm0.008$ with a maximum overlap of $59.9\pm0.3\ \%$ with BAHAMAS-0.1.

\begin{figure}
    \centering
    \includegraphics[width=\columnwidth]{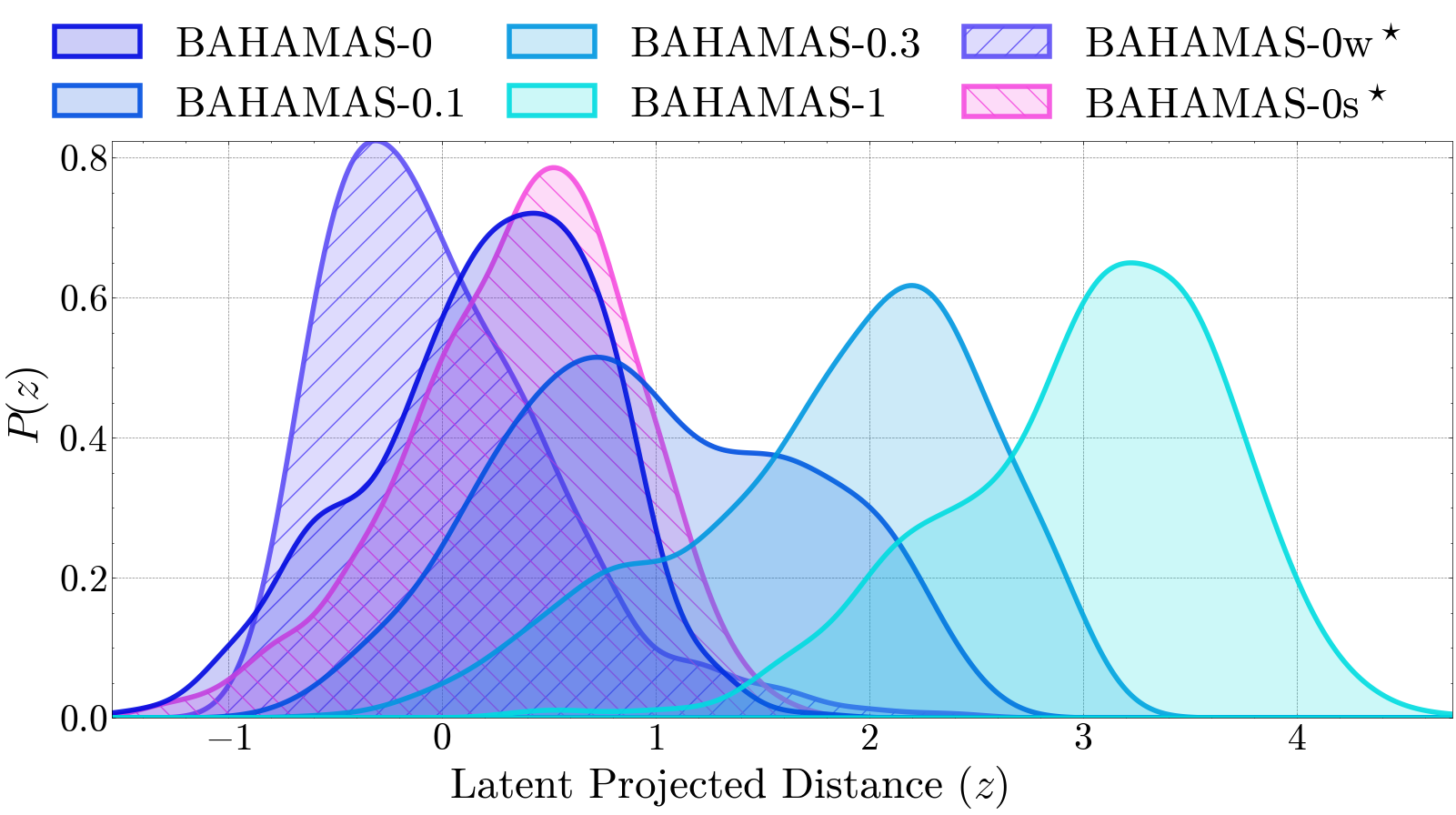}
    \caption{
7D latent space projection from our clustering algorithm into a 1D distance PDF, where each known simulation and BAHAMAS-0s is projected in the direction from the centre of their distribution to the centre of BAHAMAS-0w (blue shades and pink hatched).
        BAHAMAS-0w is projected in the direction of BAHAMAS-0 to show the greatest overlap (purple hatched).
        The distance has arbitrary units as it depends on the scale of the latent space.
        BAHAMAS-0w shares a large overlap with BAHAMAS-0, leading to the conclusion that it lies within the training domain.
    }
    \label{fig:bahamas-agn-mmd}
\end{figure}

\begin{figure*}
    \centering
    \includegraphics[width=\columnwidth]{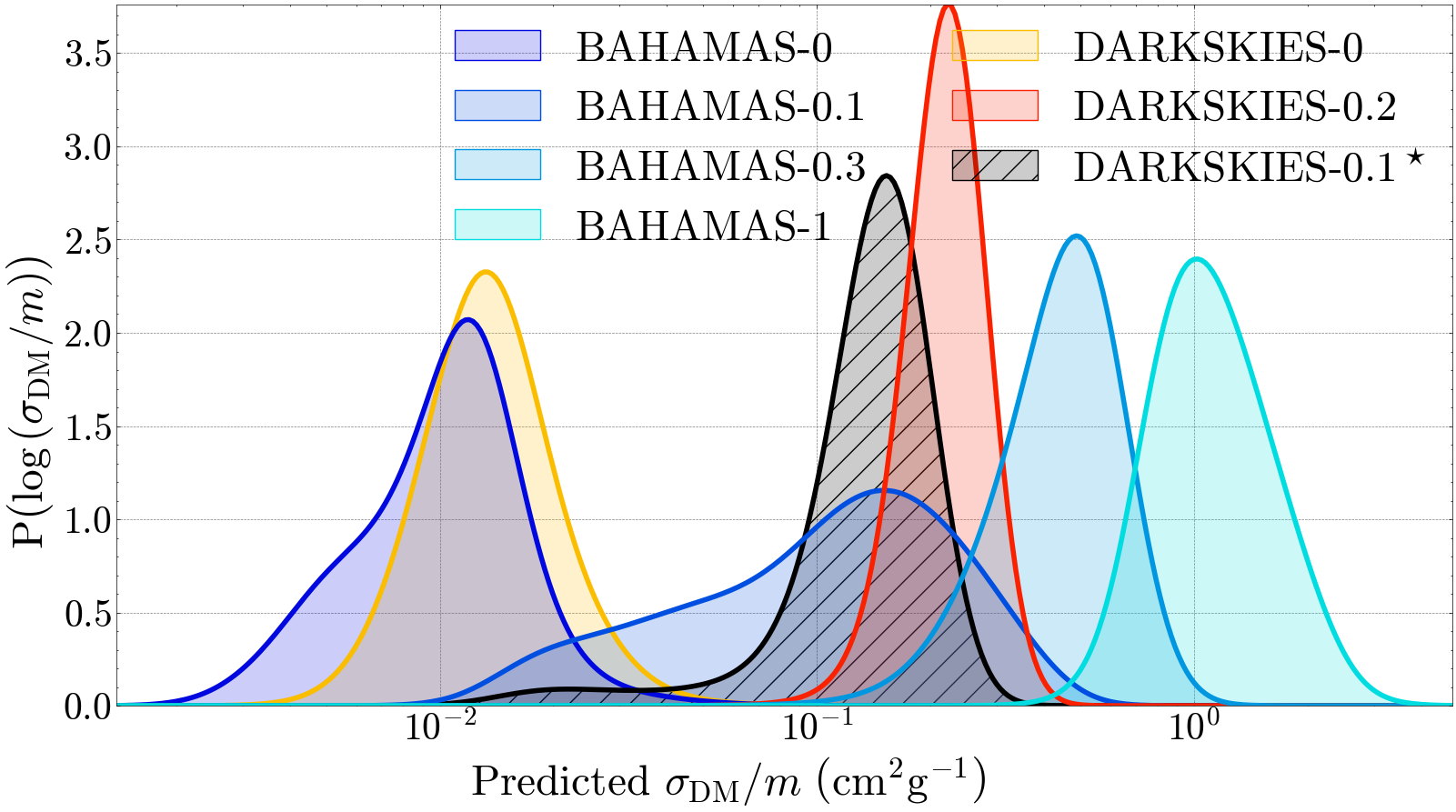}
    \includegraphics[width=\columnwidth]{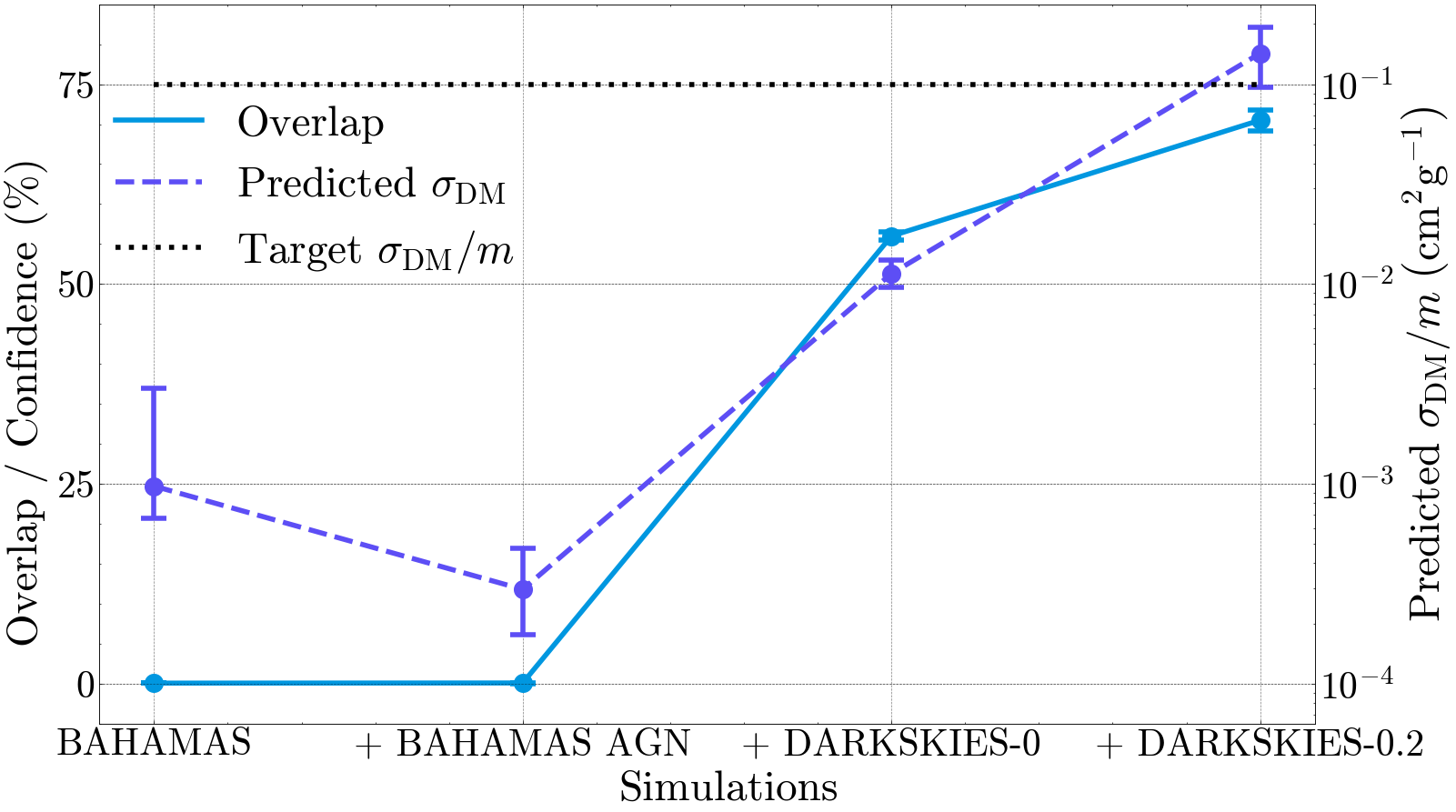}
    \caption{
        Adaptation to a new domain.
        {\it Left:} We train an ensemble of five clustering NNs with BAHAMAS, DARKSKIES-0, and DARKSKIES-0.2 as known (solid blue and red shades) and DARKSKIES-0.1 as unknown (black hatched).
        We show the combined PDF of $\log{\left(\sigma_{\rm DM}/m\right)}$ from the five runs.
        We find a cross-section of $\sim0.14\ {\rm cm^2 g^{-1}}$ for DARKSKIES-0.1, within $1\sigma$ of $0.1\ {\rm cm^2 g^{-1}}$.
        {\it Right:} Overlap confidence (left axis, solid blue), estimated $\sigma_{\rm DM}/m$ (right axis, purple dashed), and target $\sigma_{\rm DM}/m$ (right axis, black dotted) for DARKSKIES-0.1 as we progressively add known simulations during training.
        We start with BAHAMAS-0 and BAHAMAS-SIDM, then add the BAHAMAS-AGN simulations, DARKSKIES-0, and finally DARKSKIES-0.2.
        We see that the overlap confidence closely follows the convergence of DARKSKIES-0.1 onto the target $\sigma_{\rm DM}/m$ with an overlap of 70\% resulting in DARKSKIES-0.1 being within $1\sigma$ of the target $\sigma_{\rm DM}/m$.
    }
    \label{fig:darkskies-0.1}
\end{figure*}

Next, we introduce additional CDM simulations, BAHAMAS-0w and BAHAMAS-0s, which differ in their level of AGN feedback.
All the previous BAHAMAS simulations are now known, while the extra BAHAMAS-CDM simulations are given unique unknown labels.
Figure~\ref{fig:bahamas-agn-mmd} shows the distributions of each simulation in the direction towards BAHAMAS-0w.
We show the projection of the BAHAMAS-0w distribution along the direction towards BAHAMAS-0, which exhibits the greatest overlap with a known class.
An analogous figure can be produced for BAHAMAS-0s, which produces qualitatively similar results.
The estimated $\sigma_{\rm DM}/m$ for BAHAMAS-0w and BAHAMAS-0s are $\sigma_{\rm DM}/m={6.7^{+12.1}_{-4.4}}\times10^{-3}\ {\rm cm^2g^{-1}}$ and $\sigma_{\rm DM}/m={12.5^{+16.1}_{-8.1}}\times10^{-3}\ {\rm cm^2g^{-1}}$, respectively.
For BAHAMAS-0w, the maximum overlap is $83.9\pm0.6\ \%$ with BAHAMAS-0, while for BAHAMAS-0s, the maximum overlap is $90.1\pm1.3\ \%$ with BAHAMAS-0.
The overlap between the two AGN simulations is $77.1\pm0.7\ \%$.

\subsection{Realistic Out-of-Domain Data}

Finally, we want to observe the effect of choosing a realistic dataset that is initially outside the training domain, and how adding additional simulations progressively brings the test set into the training domain.
To do this we set the unknown dataset as DARKSKIES-0.1 throughout.
We initially train three models over 150 epochs on the fiducial BAHAMAS set, with DARKSKIES-0.1 as the unknown dataset.
We estimate the cross-section (where the truth is $\sigma_{\rm DM}=0.1\ {\rm cm^2 g^{-1}}$), project the latent space, and calculate the overlap integral with the most similar simulation.
We then rerun the training, with each time including more simulations: BAHAMAS + AGN, then DARKSKIES-0, and finally DARKSKIES-0.2.
In the case of DARKSKIES-0 we are forced to give it a value for $\sigma_{\rm DM}/m$; therefore, we motivate our selection for CDM simulations to be in line with the effective $\sigma_{\rm DM}/m$ found in \cite{harvey2019observable}.
Although DARKSKIES is significantly higher resolution than BAHAMAS, we do not want the model to distinguish differences within CDM simulations.
In fact, we want it to learn to align all CDM simulations, so we assign DARKSKIES-0 the same classification label as BAHAMAS-0.
Since this $\sigma_{\rm DM}/m$ represents the upper limit of the model's ability to distinguish SIDM from CDM, this assignment should not introduce any bias.

The left-hand panel of Figure~\ref{fig:darkskies-0.1} shows the combined probability distributions of $\sigma_{\rm DM}/m$ for each simulation in the final trained NN including all datasets.
Blue shades correspond to BAHAMAS simulations and red shades to DARKSKIES, with the unknown test DARKSKIES-0.1 represented by black hatching.
The BAHAMAS simulations are estimated to have $\sigma_{\rm DM}/m$ of $(1.04^{+0.54}_{-0.48})\cdot10^{-2}\ {\rm cm^2 g^{-1}}$, $0.113^{+0.119}_{-0.076}\ {\rm cm^2 g^{-1}}$, $0.444^{+0.185}_{-0.151}\ {\rm cm^2 g^{-1}}$, and $1.07^{+0.554}_{-0.315}\ {\rm cm^2 g^{-1}}$ for BAHAMAS-0, BAHAMAS-0.1, BAHAMAS-0.3, and BAHAMAS-1 respectively.
For the DARKSKIES simulations, the estimations are $(1.31^{+0.68}_{-0.43})\cdot10^{-2}\ {\rm cm^2 g^{-1}}$ and $0.217^{+0.056}_{-0.049}\ {\rm cm^2 g^{-1}}$ for DARKSKIES-0 and DARKSKIES-0.2 respectively, and for the unknown test DARKSKIES-0.1, the NN estimates $0.143^{+0.050}_{-0.046}\ {\rm cm^2 g^{-1}}$.

The right-hand panel of Figure~\ref{fig:darkskies-0.1} shows the largest overlap between DARKSKIES-0.1 and the known simulations, on the right-hand y-axis, along with the estimated $\sigma_{\rm DM}/m$ on the left-hand y-axis, for each set of training simulations.
As the overlap increases, the estimation converges towards the correct cross-section, with an overlap of $70.5\pm1.3\ \%$ resulting in an estimate within 1$\sigma$ of the target value.
DARKSKIES-0.1 has zero overlap with the BAHAMAS simulations, which could be the reason for the large error when DARKSKIES-0.2 is not included as BAHAMAS is not able to provide sufficient upper limit support.

\section{Discussion}\label{sec:dicussion}

We have presented a semi-supervised clustering algorithm that provides both confidence estimation and interpretability for constraints on the DM self-interaction cross-section.
We demonstrate that the model can identify additional physical features in the dataset, detect out-of-domain data, and constrain idealised observations in a $\sigma_{\rm DM}/m=0.1\ {\rm cm^2 g^{-1}}$ universe.
However, several aspects remain that require further discussion and development.

\subsection{Interpreting the Self-Organising Latent Space}

We first show in Figure~\ref{fig:bahamas-agn-pca} that clustering in a high-dimensional latent space allows us to analyse the model's learned features, using either an information ordered bottleneck layer or PCA.
We can directly observe how the model is learning different levels of baryonic feedback compared to $\sigma_{\rm DM}/m$.
However, without methods to identify how the NN is learning or without a well sampled latent space, interpreting what physical properties each latent dimension represents can be challenging and uncertain.
This is often due to the fact that physical features do not extend linearly along individual latent dimensions; rather, each latent feature may correspond to a non-linear combination of the simulation properties.
In the case of Figure \ref{fig:bahamas-agn-pca}, where AGN feedback and $\sigma_{\rm DM}/m$ were the only varying parameters, the latent features were relatively straightforward to interpret.

To improve interpretability, additional simulations spanning a wider range of parameter space are needed to cross-correlate physical properties with latent features.
It may also be beneficial to use methods that can directly identify which input features the NN uses to inform its estimations, such as feature importance, activation maps, counterfactual examples, or generative models.

Another avenue for improvement is how the network learns to cluster data.
Currently, latent vector similarity is minimised or boosted if they share the same class; however, using a class-level similarity metric, including physical properties, such as self-interaction cross-section; summary statistics, such as density profiles; or features extracted from NNs, this could provide a more robust method for clustering based on feature similarity.

\subsection{Confidence Estimation}

We show that high-dimensional latent spaces provide a natural means to measuring similarity and detect out-of-domain datasets.
We can use different metrics to measure similarity, such as overlap, Earth Mover Distance (EMD), KS-test, and classification accuracy.
Each metric has its own advantages and limitations.
Overlap, EMD, and the KS-test require, in the simplest form, 1D projects, while classification accuracy operates in the full latent space and arises naturally from our architecture, albeit as an approximation.
Classification accuracy and overlap are the most interpretable, while EMD is the most sensitive at the extremes of very high or very low similarity.
Ultimately, we adopt overlap as our primary metric due to its simplicity, interpretability, and effectiveness as a proxy for similarity.
However, future work will explore more robust metrics, including higher-dimensional overlap and methods that account for similarity across multiple reference datasets.

\subsection{Application to Real Data and Reducing Uncertainties}

Our ultimate goal is to constrain $\sigma_{\rm DM}/m$ in our universe; therefore, before we can pass observations to our NN, we must first forward model our total mass maps into realistic weak lensing convergence or shear maps, incorporating instrumental effects, Fourier boundary artifacts from the shear-to-convergence transformation, and statistical noise.
This degradation will inevitably increase the uncertainty in our estimated $\sigma_{\rm DM}/m$ values and reduce our constraining power.
Future work will aim to reduce these uncertainties by:

\begin{itemize}
    \item Adding more simulations to improve generalisation, especially in new domains
    \item Use higher-resolution simulations with greater parameter space coverage
    \item Train for more epochs and train more NNs for improved ensemble learning
    \item Improve the NN architecture with features from models such as vision transformers or ConvNeXt
\end{itemize}

Another important consideration is how we handle the non-physical value of $\sigma_{\rm DM}/m$ assigned to CDM.
While SIDM simulations follow $\sigma_{\rm DM}/m\gg0.01\ {\rm cm^2 g^{-1}}$, the arbitrary CDM $\sigma_{\rm DM}/m$ does not currently cause significant bias; however, future simulations approaching this threshold may require us to exclude low-resolution CDM simulations if they introduce confusion with SIDM models.

\subsection{Comparison with Simulation Based Inference Methods}

Simulation-based inference (SBI) is a powerful approach for learning posteriors or likelihoods directly from simulations, providing uncertainty estimates for individual observations.
However, SBI typically operates on per-sample posteriors, whereas in cosmology, macroscopic parameters, such as the dark matter self-interaction cross-section, are generally constrained through observational ensembles.
While our method shares many similarities with traditional SBI through the use of simulations to obtain posteriors for parameters, it differs by aggregating predictions from many observations to recover a posterior for a macroscopic parameter, even in sparsely sampled domains where traditional machine learning methods, including SBI, struggle to interpolate.

\subsection{Domain Variance and Adaptation}

It is clear from Figure \ref{fig:darkskies-0.1} that the clustering algorithm in its current state does not generalise well beyond that of its training domain, unable to predict the cross-section in the DARKSKIES suite of simulations, a common problem in ML.
In the scope of this work we required an algorithm that did not generalise well in order to evidence how it worked (should it have been perfectly generalised, the right-hand side of Figure \ref{fig:darkskies-0.1} would be flat).
However, even in the case where there was perfect generalisation we would still require empirical insight when applied to real observations to inform us of the model's confidence in its estimates.
Therefore, our method and domain adaptation are complementary and both will need to be combined before we apply this to data to identify if the domain adaption is correctly aligning the features in the two domains, which we will show in future work.

\section{Conclusions}
\label{sec:conc}

We have presented a machine learning method to measure the dark matter self-interaction cross-section, $\sigma_{\rm DM}/m$, from two dimensional gravitational lensing mass maps that can return robust confidence limits in its estimates.
Since we cannot observe multiple universes with different $\sigma_{\rm DM}/m$, we must rely on simulations for inference.
However, simulations do not perfectly replicate observations, which presents a challenge for simulation-based inference.
We have developed a clustering neural network (NN) architecture capable of recovering macroscopic parameters in a sparsely sampled parameter space while producing an interpretable latent space that provides a metric for assessing the confidence of estimations on unknown datasets.
The NN learns features from both known and unknown datasets during training, enabling it to cluster samples based on their similarities in a high-dimensional latent space.

Through a series of targeted tests, we show that when observations resemble the training domain, our method can recover $\sigma_{\rm DM}/m$ within $1\sigma$, even with limited simulated coverage of the parameter space.
The high-dimensional latent space also enables exploration of secondary physical features, such as variations in black-hole feedback, even when these are not explicitly labelled during training.
Furthermore, by expressing the latent space in higher dimensions and measuring the projected overlap between different domains, we can quantify estimation confidence.
This estimation confidence allows us to determine whether a dataset lies within the training domain, suggesting shared features and reliable interpolation, or is out-of-domain, indicating the dataset is foreign with potential extrapolation and unreliable estimations.
We demonstrate that when presented with a dataset of uniform noise, the network correctly identifies it as out-of-domain relative to the BAHAMAS simulations.
Finally, we show that if our observations, DARKSKIES-0.1, belong to a different suite of simulations without prior exposure, the NN fails to recover the target $\sigma_{\rm DM}/m$.
However, by leveraging higher dimensions and similarity metrics, we show that NNs can still be used in trustworthy ways for scientific inference.

Our next steps will be to move towards applying this architecture to real observations to obtain a constraint competitive with traditional methods for $\sigma_{\rm DM}/m$.
As such, it will be essential to generate realistic mock observations from simulations, incorporating instrumental noise and observational effects.
We also aim to train on a broader range of cosmological simulations to better sample the latent space and marginalise over non-physical differences across simulation suites.
For future improvements, we aim to incorporate domain adaptation techniques to reduce the performance gap on datasets that differ from the training domain.
Finally, we can explore methods for NN interpretability to gain deeper insight into what the NN is learning, further improving trust of machine learning in science. 

The method we have presented here represents a blueprint for the use of machine learning in scientific inference.
The flexibility of compact clustering to manipulate the learned features amongst a variety of different simulation suites enables us to build an enriched latent-space, that incorporates a host of different simulations.
This way we can numerically marginalise over all-unknowns delivering robust and trust-worthy inference of cosmological parameters.

\section*{Acknowledgements}

This work was supported by the Swiss State Secretariat for Education, Research and Innovation (SERI) under contract number 521107294.
We would like to thank Andrew Robertson and Ian McCarthy for developing the BAHAMAS-SIDM simulations.

%%%%%%%%%%%%%%%%%%%%%%%%%%%%%%%%%%%%%%%%%%%%%%%%%%
\section*{Data Availability}

Data from \cite{mccarthy2016bahamas,robertson2019observable,harvey2025darkskies} for the BAHAMAS and DARKSKIES simulations are available on request from the original authors.
Access to the code and trained weights of the neural networks can be found via \url{https://github.com/EthanTreg/Bayesian-DARKSKIES}.

%%%%%%%%%%%%%%%%%%%% REFERENCES %%%%%%%%%%%%%%%%%%

\bibliographystyle{aa}
\bibliography{physicalised_cluster,bibliography_drh}

%%%%%%%%%%%%%%%%%%%%%%%%%%%%%%%%%%%%%%%%%%%%%%%%%%

%%%%%%%%%%%%%%%%% APPENDICES %%%%%%%%%%%%%%%%%%%%%

\appendix

\section{Compact Clustering via Label Propagation}
\label{app:cclp}

In Section~\ref{sec:cluster} we outline the three loss functions used in our method, with Equation~\ref{eq:cclp_loss} introduced by \cite{kamnitsas2018semi}.
In this section, we derive the matrices $\mat{T}\in\mathbb{R}^{N\times N}$ and $\mat{H}^{(s)}\in\mathbb{R}^{N\times N}$, latent vectors $\mat{Z}\in\mathbb{R}^{N\times\left|\mathcal{Z}\right|}$, and a set of class probabilities, $\mat{Y}'\in\mathbb{R}^{N\times C}$, for the set of classes, $C$, with batch size $N$ and latent space $\mathcal{Z}$.
A portion of the dataset is unlabelled, $N=N_L+N_U$, with latent vectors $\mat{Z}_L$ and $\mat{Z}_U$, where $L$ and $U$ subscripts correspond to the labelled and unlabelled samples, respectively.
As we want to find the transition and target transition matrices, $\mat{H}$ and $\mat{T}$, we first generate a fully connected graph in the latent space by calculating the adjacency matrix, $\mat{A}\in\mathbb{R}^{N\times N}$ from Equation~\ref{eq:adjacency}, where each element, $A_{ij}$, is the weight of an edge in the graph and represents the similarity between samples $i$ and $j$.

\begin{equation}
    \label{eq:adjacency}
    \mat{A} = \exp{\left(\mat{Z}\mat{Z}^{\rm T}\right)}
\end{equation}

We then row-wise normalise $\mat{A}$, Equation~\ref{eq:transition}, so that each element of the transition matrix, $H_{ij}$, is the probability of a transition from sample $i$ to sample $j$.

\begin{equation}
    \label{eq:transition}
    H_{ij} = \frac{A_{ij}}{\sum_k{A_{ik}}}
\end{equation}

We structure $\mat{H}$ so that the labelled and unlabelled samples are arranged as Equation~\ref{eq:transition-order}.

\begin{equation}
    \label{eq:transition-order}
    \mat{H} =
    \left[\begin{matrix}
        \mat{H}_{LL} & \mat{H}_{UL} \\
        \mat{H}_{LU} & \mat{H}_{UU}
    \end{matrix}\right]
\end{equation}

From $\mat{H}$, we can propagate labels from the known samples to the unknown samples based on their probability of transition, forming the class posterior matrix $\mat{\Phi}=\left[\begin{matrix}\mat{Y}_L\\\mat{\Phi}_U\end{matrix}\right]\in\mathbb{R}^{N\times C}$, where $\mat{Y}_L\in\mathbb{R}^{N_L\times C}$ are the one hot vectors of the known class labels and $\mat{\Phi}_U\in\mathbb{R}^{N_U\times C}$ is the unlabelled class posteriors given by Equation~\ref{eq:posteriors}.

\begin{equation}
    \label{eq:posteriors}
    \mat{\Phi}_U = \left(\mat{I} - \mat{H}_{UU}\right)^{-1} \mat{H}_{UL} \mat{Y}_L
\end{equation}

Then, from $\mat{\Phi}$, we can calculate $\mat{T}$, which acts as a soft target probability for a transition between two class labels, maximising same class transitions, while minimising inter-class transitions, given by Equation~\ref{eq:target-transition}.

\begin{equation}
    \label{eq:target-transition}
    T_{ij} = \sum_{c=1}^C{\frac{\phi_{ic} \phi_{jc}}{\sum_{k=1}^N{\phi_{kc}}}}
\end{equation}

Finally, we want to minimise the cross-entropy between $\mat{T}$ and $\mat{H}$; however, we can form Markov chains to enable class propagation along high density regions, preserving the structure of the graph.
Therefore, the probability that a Markov process starts from sample $i$, performs $\left(s-1\right)$ steps within the same class, and then transitions to sample $j$ is given by Equation~\ref{eq:markov-chain}.

\begin{equation}
    \label{eq:markov-chain}
    \mat{H}^{\left(s\right)} = \left(\mat{H}\circ\mat{M}\right)^{s-1}\mat{H}
\end{equation}

Where $\circ$ is the Hadamard product and $\mat{M}=\mat{\Phi}\mat{\Phi}^{\rm T}\in\mathbb{R}^{N\times N}$ is the probability that two samples belong to the same class, with $H_{ij}M_{ij}$ being the approximate joint probability of transitioning from sample $i$ to sample $j$ and the two samples belonging to the same class.
Therefore, we can construct $\mathcal{L}_{\rm CCLP}$ as the cross-entropy between the target transition, $\mat{T}$, and the transition probability along the Markov chain, $\mat{H}^{\left(s\right)}$ as shown in Equation~\ref{eq:cclp_loss}.
For more details see \cite{kamnitsas2018semi}.

\section{Network Architecture}
\label{app:architecture}

\begin{figure}
    \centering
    \includegraphics[width=\columnwidth]{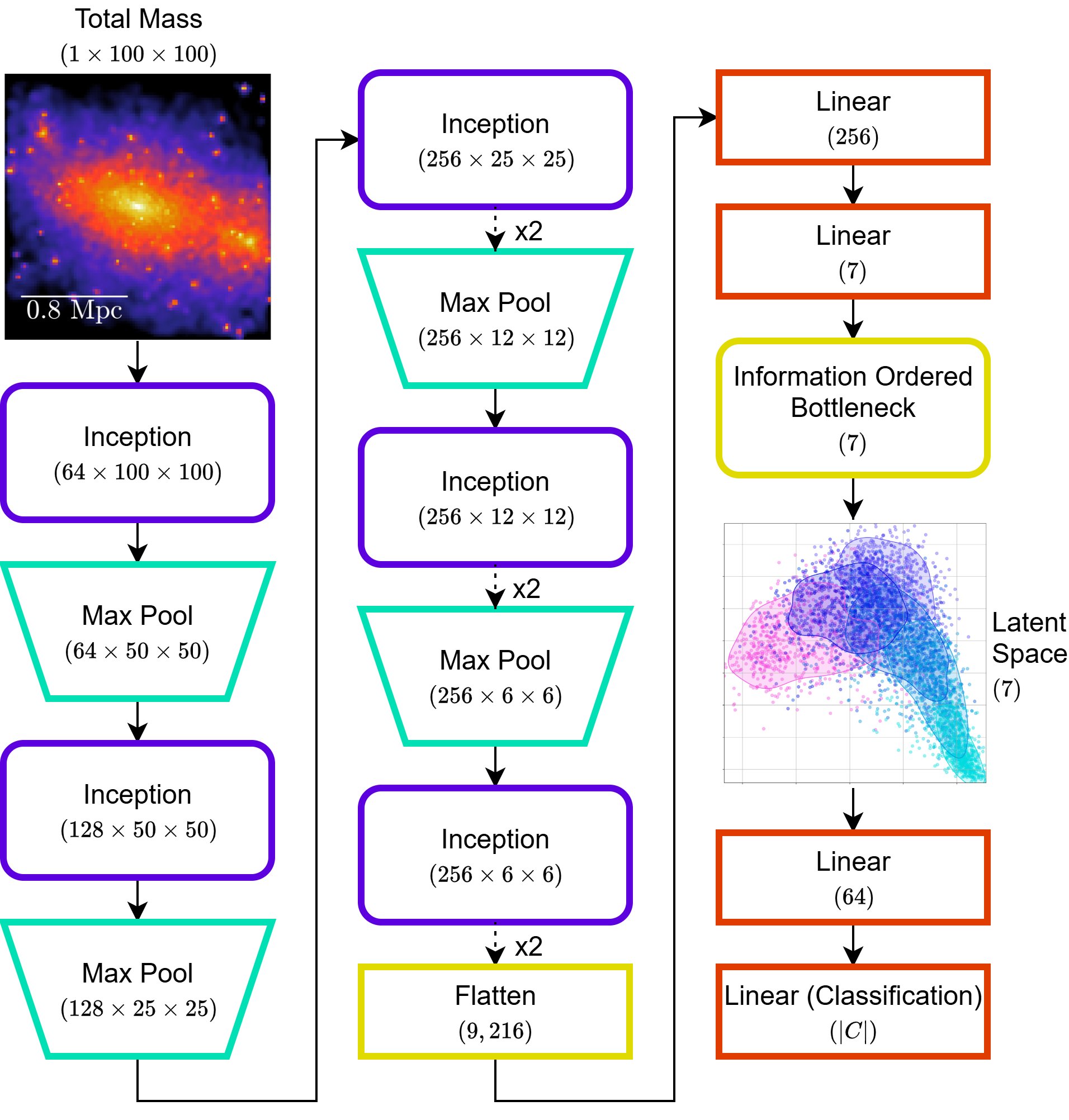}
    \caption{
        Full network architecture used in the paper.
        The network is composed of inception blocks (purple, rounded rectangles), show in Figure~\ref{fig:inception-block}, max pooling layers (turquoise, trapeziums), linear layers (red, rectangles), a flatten layer (yellow, rectangle), and an information ordered bottleneck layer (yellow, rounded rectangle).
        The output dimensions of each layer is represented underneath the name of the layer, the arrows represent the direction of data flow and if there is a $\times2$ next to the arrow, then the layer is repeated twice.
        The input is the total mass maps and optionally X-ray maps of galaxy clusters and the latent space is where we calculate the loss functions from equations~\ref{eq:cclp_loss} and~\ref{eq:dist_loss} as well as our feature analysis.
    }
    \label{fig:network-architecture}
\end{figure}

\begin{figure}
    \centering
    \includegraphics[width=\columnwidth]{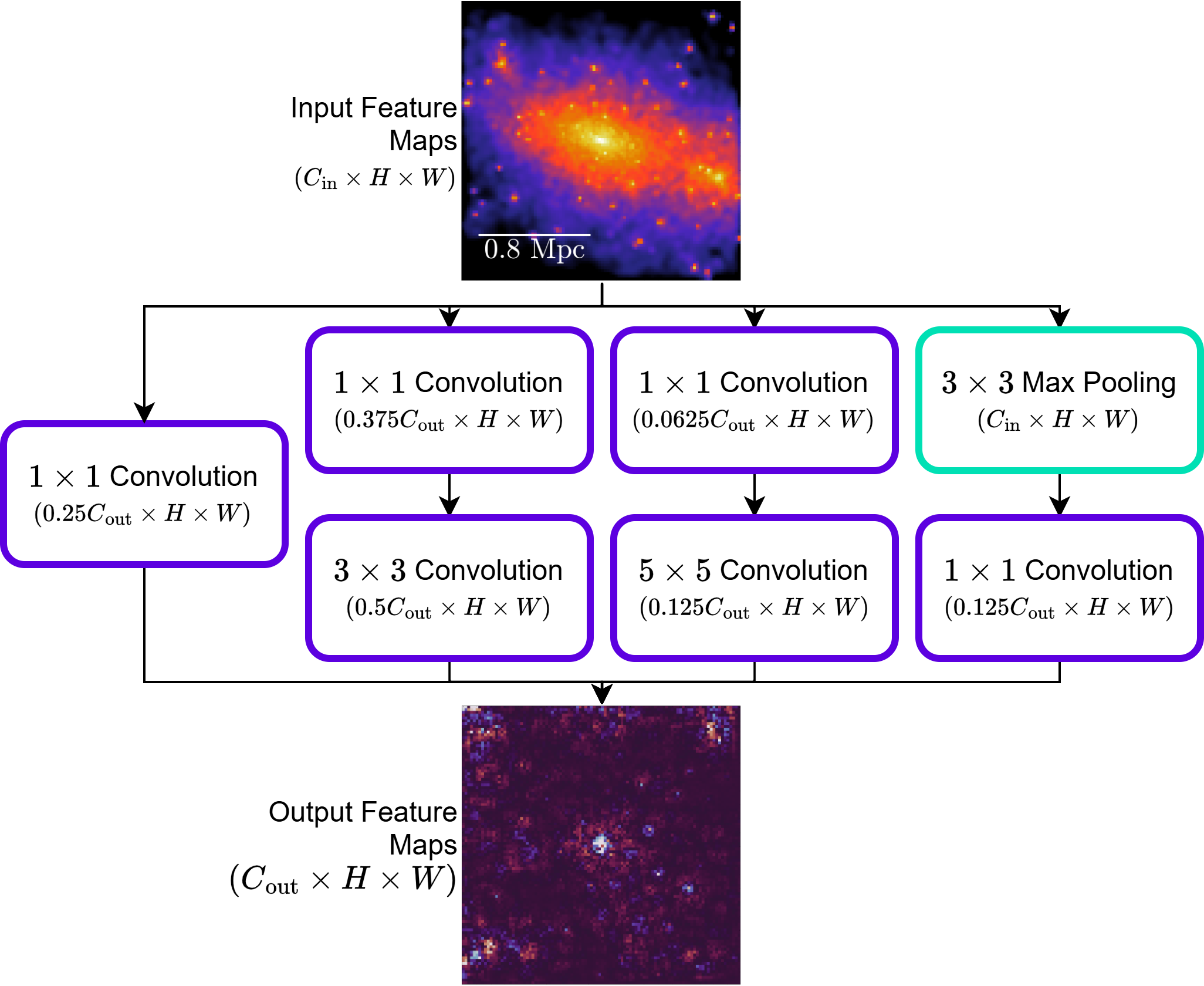}
    \caption{
        Inception block used in our full network architecture, Figure~\ref{fig:network-architecture}.
        The block is composed of convolutional layers (purple) and a max pooling layer (turquoise).
        The inception block uses a mix of $1\times1$, $3\times3$, and $5\times5$ convolution, all with same padding and a stride of one to maintain the input dimensions $H\times W$.
        We show the number of channels as a fraction of the output number of channels as this is variable, as shown in Figure~\ref{fig:network-architecture}.
    }
    \label{fig:inception-block}
\end{figure}

We show in Figure~\ref{fig:network-architecture} the full architecture used in this paper.
As mentioned in Section~\ref{sec:architecture}, the input is the total mass maps and optionally X-ray maps and the output is a 7D latent space and classification scores.
The network is composed of inception blocks (purple, rounded rectangle), Figure~\ref{fig:inception-block}, max pooling (turquoise, trapeziums), and linear layers (red, rectangles).
Some inception blocks are followed by a $\times2$ representing that this layer is repeated twice and the output shape from each layer is shown under the layer name with the last layer depending on the number of classes, $\left|C\right|$, used during training, which includes the number of both known and unknown simulations with unique $\sigma_{\rm DM}/m$.
The inception blocks use convolutional layers with different kernel sizes to learn spatial features within different receptive fields.
The max pooling downscales the input with inception blocks following as this enables the convolutional layers within the inception blocks to learn features on different scales as each downscale increases the receptive field of the convolutional kernel.
The flatten followed by linear layers converts the spatial domain to a regression and classification domain with a linear layer outputting seven features followed by an information ordered bottleneck creating the latent space and two more linear layers after the latent space for the classification loss function.
The information ordered bottleneck aims to order the latent dimensions based on loss function importance, so the first dimensions should contribute the most to minimising the loss, and therefore, containing the most information, Figure~\ref{fig:latent-space} shows an example of a 7D latent space.
There are many more advanced network architectures used in computer vision; however, we found worse performance when using Inception-v4 and ConvNeXt compared to our architecture, likely due to the large downscaling in the stem of these networks and the importance of small scale features in our data, resulting in reduced sensitivity to the small scale features; however, future work will investigate incorporating the advancements of these new architectures to improve the performance of our results.

The inception block show in Figure~\ref{fig:inception-block} uses several convolutional layers (purple) with different kernel sizes and a max pooling layer (turquoise) to extract features.
The input to the inception block is the output from the previous layer in the network, or the input to the network, with the output from each layer shown under the layer name.
The height and width of the input features remains unchanged, while the number of channels is a fraction of the output number of channels for that inception block.
$1\times1$ convolutional layers help with reducing the number of channels before $3\times3$ and $5\times5$ convolutional layers, reducing the computational cost, while a mix of max pooling and different sized convolutional kernels can learn different spatial features over different receptive fields.

\section{Training Choices}
\label{sec:biases}

To evaluate the best hyperparameters for training, we perform several tests.
We used the reduced architecture as this enables faster testing and trained the NN ten times for 150 epochs for each test case to use ensemble learning, reducing training stochasticity.
We train the network on the total mass maps with the BAHAMAS simulations, with all simulations treated as known, unless otherwise stated.
Each trained NN will generate a set of estimations for each simulation, which we can treat as a posterior, then using ensemble learning, we multiply the posteriors of the ten NNs for the same test case to get a final posterior.
From the final posterior, we take the median and quantiles representing $1\sigma$ and $2\sigma$.

\begin{figure}
    \centering
    \includegraphics[width=\columnwidth]{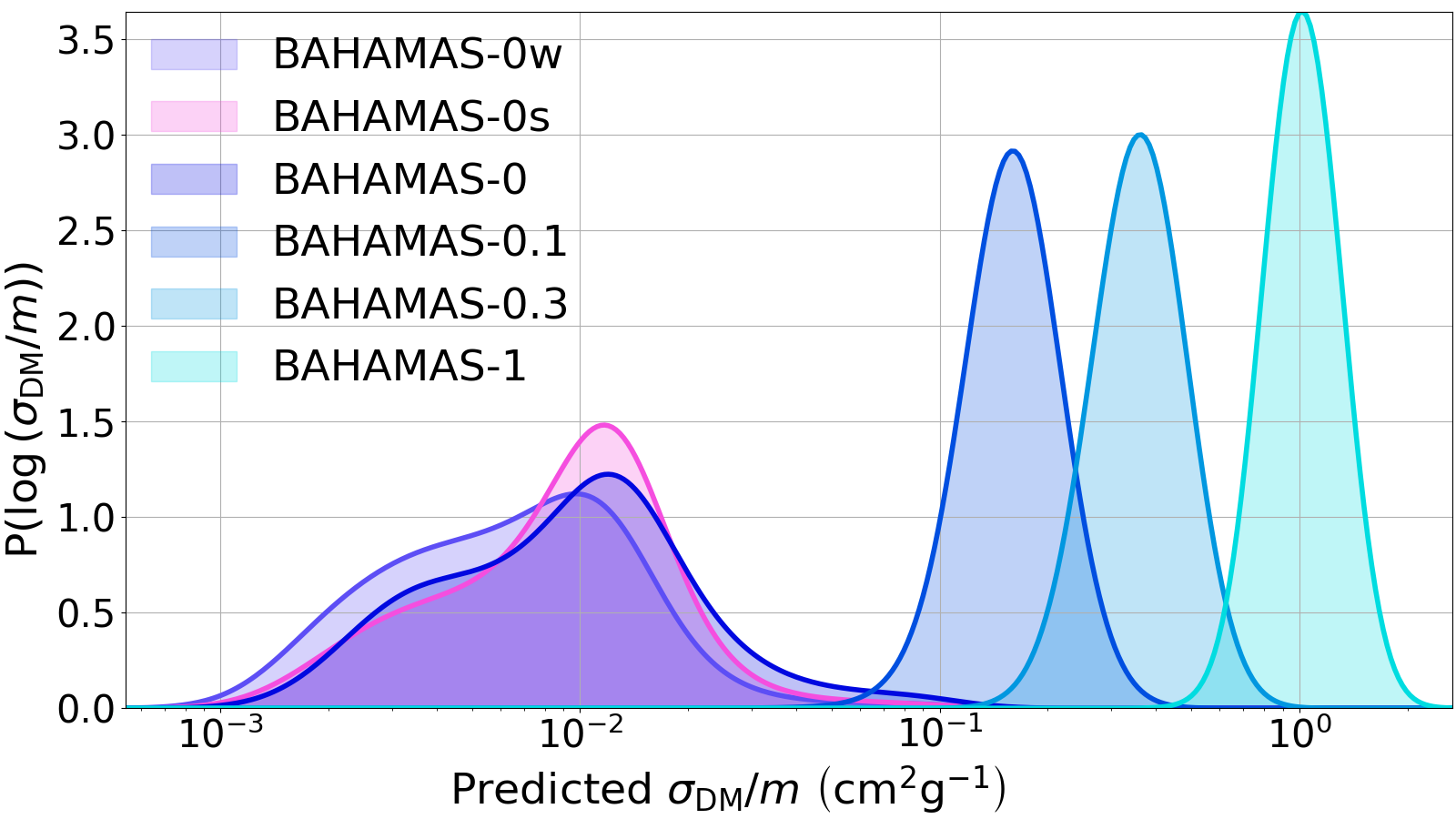}
    \caption{
        We train an ensemble of 10 clustering NNs after calibration on all BAHAMAS simulations, all treated as known.
        We show the combined PDF of $\log{\left(\sigma_{\rm DM}/m\right)}$ for the ten NNs with all simulations within $1\sigma$ of their target $\sigma_{\rm DM}/m$ except BAHAMAS-0.1 which is within $2\sigma$.
    }
    \label{fig:reduced-network}
\end{figure}

The configuration found to give the best accuracies and Gaussian like posteriors is when $\sigma_{\rm DM}/m$ is logarithmically transformed, all CDM simulations are treated as known and assigned the same effective $\sigma_{\rm}$ of BAHAMAS-CDM, and X-ray maps are excluded.
For this configurations, all BAHAMAS simulations are ${<}1\sigma$ of their target $\sigma_{\rm DM}/m$ except BAHAMAS-0.1 which is ${<}2\sigma$.
% However, FLAMINGO-0 is an order of magnitude below the target BAHAMAS effective $\sigma_{\rm DM}/m$, with an estimate of $\sigma_{\rm DM}/m=1.6^{+1.2}_{-0.6}\times10^{-3}$, within 1$\sigma$ of the FLAMINGO-0 effective $\sigma_{\rm DM}/m$.
Figure~\ref{fig:reduced-network} shows the estimated $\sigma_{\rm DM}/m$ posteriors for each simulation with all simulations having Gaussian like posteriors except BAHAMAS-CDM which are asymmetric.
The following subsections will justify each choice made.

\section{Interpreting Latent Spaces}

\begin{figure*}
    \centering
    \includegraphics[width=\textwidth]{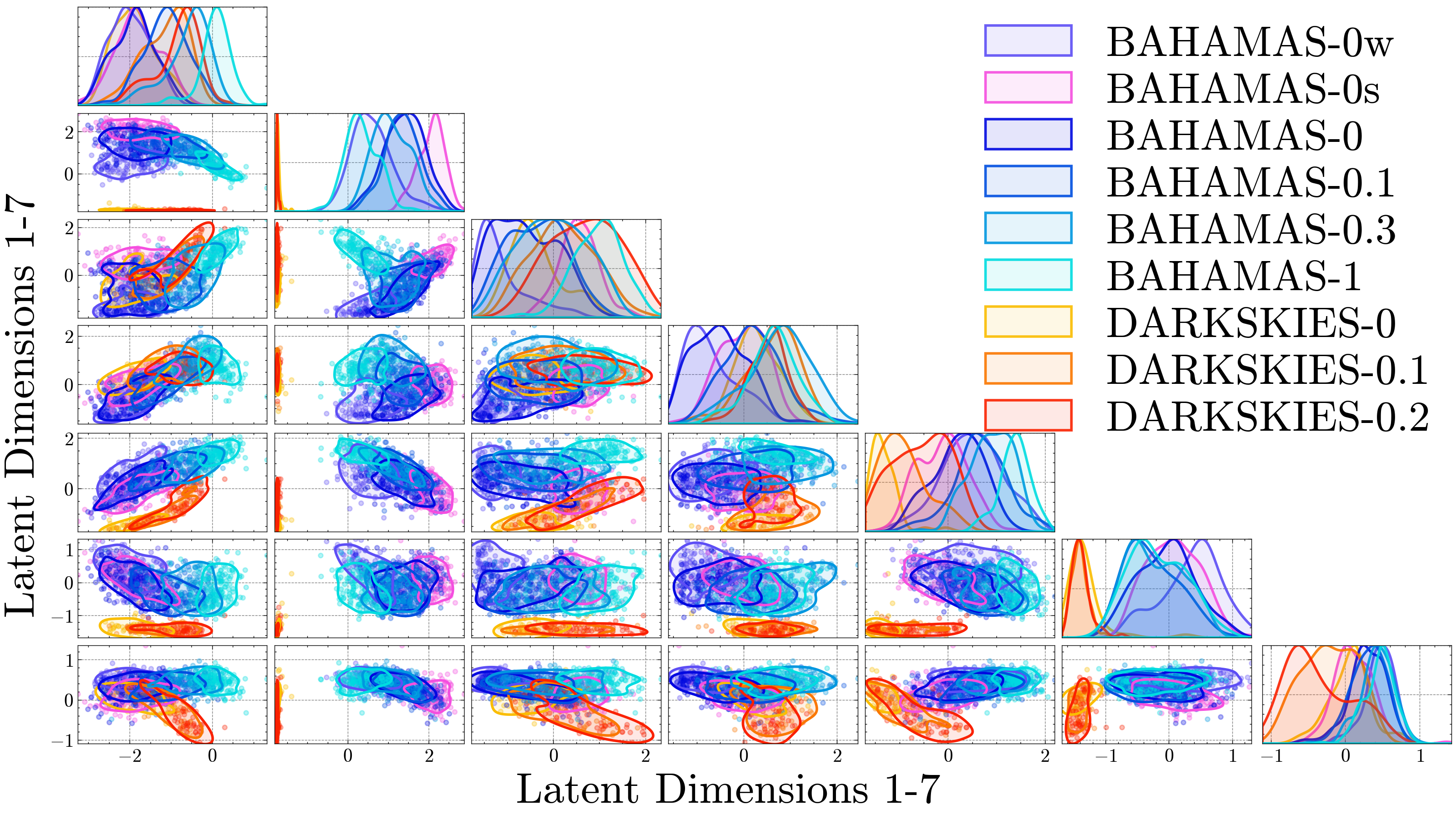}
    \caption{
        7D latent space from a NN using the full architecture trained on all BAHAMAS and DARKSKIES simulations, with all simulations treated as known and X-ray maps included.
        The diagonal subplots show the distribution of values from each simulation in each dimension, while the off-diagonal show the correlation between each dimension pair.
        The NN was trained with an information ordered bottleneck, resulting in the last dimensions (right-most columns and bottom-most rows) being biased towards containing the least important information.
        The first dimension is enforced to correspond to $\sigma_{\rm DM}/m$ via Equation~\ref{eq:dist_loss}.
        The first three dimensions show the majority of the physical information, showing $\sigma_{\rm DM}/m$, the level of AGN feedback, and the separation of BAHAMAS from DARKSKIES, while the later dimensions are either modifications of the earlier dimensions, structure dimensions to enable simulations forced apart in other dimensions to be brought closer globally, and Gaussian-like dimensions where little information is encoded.
    }
    \label{fig:latent-space}
\end{figure*}

In Figure~\ref{fig:latent-space}, we show the full 7D latent space from a NN using the full architecture trained for 150 epochs on all BAHAMAS and DARKSKIES simulations with all treated as known.
We include X-rays maps to show the more expressive latent space showing the difference between BAHAMAS-0w and BAHAMAS-0s.
The diagonals show the distribution of each simulation, normalised to a peak of one for easier visualisation, in each dimension.
The off-diagonals represent the correlations between latent dimensions for each simulation, here we can more easily see structure and how different learned features are expressed.
The left-most column and top row corresponds to the first dimension, with right increasing columns and down increasing rows representing higher dimensions.
Due to the information ordered bottleneck, as the dimension increases, it has a lower probability of carrying its information through to the loss function, and therefore, is biased towards representing less important features.

The first dimension (first column and row) is enforced to be $\sigma_{\rm DM}/m$ via the loss function from Equation~\ref{eq:dist_loss}; therefore, it is unsurprising that the simulations are ordered in increasing $\sigma_{\rm DM}/m$.
The second dimension (second column and row) primarily discriminates between DARKSKIES and BAHAMAS given by the minimal overlap between the two suites of simulations in this dimension.
However, looking at the correlation plots between dimensions one to three, we see that these hold the main physical information of $\sigma_{\rm DM}/m$ and the level of AGN feedback, identified by the T shape, most prominent in the correlation between dimensions 1-3 and 2-3, with dimension three encoding both an increasing $\sigma_{\rm DM}/m$ and level of AGN feedback.
Dimension five is the next dimension to show physical information encoded, with it being similar to the first; however with DARKSKIES offset from BAHAMAS.
Finally dimension six is the last that can be physically interpreted with it discriminating between DARKSKIES and BAHAMAS, but to a weaker extent as dimension two.
The remaining two dimensions, dimensions four and seven, do not encode any physical information.
However, four can be interpreted as a structure dimension which can allow simulations forced far away in other dimensions to be brought closer in this dimension.
Structure dimensions can help reduce opposing loss functions, such as the requirement from Equation~\ref{eq:dist_loss} for BAHAMAS-1 to be far away from BAHAMAS-0 in dimension one, while the similarity between BAHAMAS-0 and BAHAMAS-1 from Equation~\ref{eq:cclp_loss} wanting to bring them closer together as the different in $\sigma_{\rm DM}/m$ only has a minor impact on galaxy clusters, where other effects such as the level of AGN feedback or simulation resolution can contribute more to the difference.
Dimension seven, due to it being the last dimension, does not hold a lot of information and is closer to a Gaussian-like initialisation with only a minor separation between BAHAMAS and the strongest DARKSKIES.

The types of dimensions observed here are commonly reproduced by NNs trained with different combinations of data and hyperparameters, we generally see a primary dimension for each physical difference, such as $\sigma_{\rm DM}/m$, AGN feedback, and simulation suite, with secondary dimensions representing these physical differences with slight perturbations.
We also see the structure dimensions, followed by the remaining dimensions being Gaussian-like, depending on the number of latent dimensions and number of physical difference in the training data.
In the future, we will add more suites of simulations, expanding the number of physical difference and hopefully producing a more interpretable feature space, while also investigating more quantitive methods for identifying what each latent dimension is learning.

%%%%%%%%%%%%%%%%%%%%%%%%%%%%%%%%%%%%%%%%%%%%%%%%%%

% Don't change these lines
% \bsp	% typesetting comment
\label{lastpage}

\end{document}